\documentclass[12pt,journal,a4paper,draftcls,oneside,onecolumn]{IEEEtran}
\usepackage{times} 
\usepackage{graphicx}
\usepackage{cite}
\usepackage{citesort}
\usepackage{cuted}
\usepackage{makecell}
\usepackage{stfloats}
\usepackage{multirow}
\usepackage{color}
\usepackage{psfrag}
\usepackage{subfigure}
\usepackage{amssymb}
\usepackage{lmodern}
\usepackage{epsfig}
\usepackage{cuted}
\usepackage{stfloats}
\usepackage{pifont}
\usepackage{amsmath}
\DeclareMathOperator*{\argmax}{argmax}
\usepackage{cases}
\usepackage{array}
\usepackage{multicol}
\usepackage[T1]{fontenc}
\usepackage{comment}
\usepackage{amsthm}
\usepackage{indentfirst}
\usepackage{cases}
\usepackage[ruled, lined, linesnumbered, commentsnumbered, longend]{algorithm2e}
\SetKwInOut{KwIn}{Input}
\SetKwInOut{KwOut}{Output}
\SetKwProg{Init}{Initialization}{:}{end}
\SetKwProg{QT}{[QT Iteration]}{}{end}
\SetKwProg{DT}{[DT Iteration]}{}{end}
\SetKwProg{LT}{[Bisection Iteration of $\eta_l$]}{}{end}
\graphicspath{{figures/}}
\theoremstyle{remark}
\newtheorem{remark}{Remark}
\makeatletter
\def\ifundefined{\@ifundefined}
\makeatother 

\begin{document}

\title{Heterogeneous Graph Neural Network for Power Allocation in
Multicarrier-Division Duplex Cell-Free Massive MIMO Systems}

\author{Bohan Li, Lie-Liang Yang, {\em Fellow,
    IEEE}, Robert G. Maunder, {\em Senior Member, IEEE}, Songlin Sun, {\em Senior Member, IEEE}, Pei Xiao, {\em Senior Member, IEEE} \thanks{B. Li, L.-L. Yang and R. Maunder are with the School of Electronics
    and Computer Science, University of Southampton, SO17 1BJ,
    UK. (E-mail:
    bl2n18, lly, rm@ecs.soton.ac.uk,~http://www-mobile.ecs.soton.ac.uk/lly). S. Sun is with the School of Information and Communication Engineering, Beijing University of Posts and Telecommunications (BUPT). P. Xiao is with 5GIC \& 6GIC,
Institute for Communication Systems (ICS), University of Surrey, Guildford
GU2 7XH, UK. (Email: p.xiao@surrey.ac.uk)
  The project
    was supported in part by the EPSRC, UK, under Project EP/P034284/1 and EP/P03456X/1,
    in part by the Innovate UK project,  and in part by the China Scholarship Council
(CSC).}}

\maketitle

\begin{abstract}
In-band full duplex cell-free (CF) systems suffer from severe self-interference and cross-link interference, especially when CF systems are operated in distributed way. To this end, we propose the multicarrier-division duplex as an enabler for achieving full-duplex operation in the distributed CF massive MIMO systems, where downlink and uplink transmissions occur simultaneously in the same frequency band but on the mutually orthogonal subcarriers. To maximize the spectral-efficiency (SE), we introduce a heterogeneous graph neural network (HGNN) specific for CF systems, referred to as CF-HGNN, to optimize the power-allocation (PA). We design the adaptive node embedding layer for CF-HGNN to be scalable to the various numbers of access points (APs), mobile stations (MSs) and subcarriers. The attention mechanism of CF-HGNN enables individual AP/MS nodes to aggregate information from the interfering and communication paths with different priorities. For comparison, we propose a quadratic transform and successive convex approximation (QT-SCA) algorithm to solve the PA problem in classic way. Numerical results show that CF-HGNN is capable of achieving 99\% of the SE achievable by QT-SCA but using only $10^{-4}$ times of its operation time. CF-HGNN significantly outperforms the traditional greedy unfair method in terms of SE performance. Furthermore, CF-HGNN exhibits good scalability to the CF networks with various numbers of nodes and subcarriers, and also to the large-scale CF networks when assisted by user clustering.

%In-band full-duplex cell-free (CF) systems suffer from severe self-interference and cross-link interference, especially when CF systems are operated in distributed way. To this end, we introduce the multicarrier-division duplex as an enabler for achieving full-duplex operation in the distributed CF massive MIMO systems, where downlink and uplink transmissions occur simultaneously in the same frequency band but on the mutually orthogonal subcarriers. To maximize spectral efficiency (SE), we introduce a heterogeneous graph neural network (HGNN) specific for CF systems, referred to as CF-HGNN, to optimize the power-allocation (PA). We design the adaptive node embedding layer for CF-HGNN to be scalable to the various numbers of access points, mobile stations and subcarriers. For comparison, we propose a quadratic transform and successive convex approximation (QT-SCA) algorithm to solve the PA problem in classic way. Numerical results show that CF-HGNN is capable of achieving 99\% of the SE achievable by QT-SCA but using only $10^{-4}$ times of its operation time. CF-HGNN can  significantly outperform the traditional greedy unfair method in terms of the SE performance. Furthermore, CF-HGNN exhibits good scalability to the CF networks with various numbers of nodes and subcarriers, and also to the large-scale CF networks when assisted by user clustering.
\end{abstract}

\begin{IEEEkeywords}
Multicarrier-division duplex, cell-free, massive MIMO, deep learning, heterogeneous graph neural network, power allocation, spectral efficiency.
\end{IEEEkeywords}

\IEEEpeerreviewmaketitle

\section{Introduction}\label{section:MDDGNN:intro}
With the unprecedented increase of data-hungry devices, wireless communication communities have been continuing to thrive in the last few of years to seek the cutting-edge techniques for further improving spectral efficiency (SE) and quality of services (QoSs). In this regard, cell-free massive MIMO (CF-mMIMO) and in-band full-duplex (IBFD) have been envisioned as the main pillars of future wireless communication systems. Specifically, by integrating the concepts of boundaryless cells and distributed antenna mMIMO system, CF-mMIMO is capable of providing the users within a coverage area with not only higher data rates but also more consistent service than the conventional small-cell or co-located mMIMO systems \cite{demir2021foundations}. On the other side, IBFD has the potential to double the SE with off-the-shelf system resource and hence, can resolve the critical defects of the traditional half-duplex (HD) modes, such as the inefficient usage of time-frequency resource and the high latency generated by both physical and network layer operations~\cite{goyal2015full}. 

%CF power很多， CF FD很少，原因两个，自干扰和互干扰； PA 很复杂
However, to date, although there exist a large number of works on CF-mMIMO \cite{elhoushy2021cell} and IBFD systems \cite{kolodziej2019band}, the combination of these two techniques has rarely been addressed. One of the underlying reasons is that IBFD-based CF (IBFD-CF) systems suffer from the problems of severe self-interference (SI) and cross-link interference (CLI). Here, SI is referred to the interference imposed by the transmitting signals of access point (AP) or mobile station (MS) on its receiving signals. SI may be mitigated by propagation-, analog- and digital-domain self-interference cancellation (SIC) methods \cite{everett2014passive,quan2017two,ahmed2015all,liang2015digital,ghoraishi2015subband}. However, the implementation of SIC imposes large complexity and resource burden on the small-scale APs in CF networks \cite{li2021multicarrier}. Moreover, none of the existing methods can fully suppress SI, and unavoidably, the residual SI causes the performance degradation of desired transmission. CLI, known as the inter-AP interference (IAI) or inter-MS interference (IMI), has been shown to be the most serious problem in multi-cell IBFD and dynamic time-division duplex (TDD) systems \cite{da2021full}. The situation is even exacerbated in the CF networks, where the IAI received by an AP may come from various directions. Noticeably, compared with IAI, IMI is more difficult to manage, as the baseband processors of MSs are usually not as powerful as those of APs. Fortunately, the transmit power of MSs is relatively low and in this case, the IMI can be greatly mitigated in the analog domain using the power control and/or user scheduling approaches \cite{kim2020dynamic}.
 
%Many approaches have been presented to efficiently mitigate CLI, such as beamforming suppression \cite{huang2018reaping,de2018distributed}, but they can not be directly applied to IBFD-CF systems, since APs employed with small number of antennas are not able to suppress interference from various directions, not mention to the much tiny MS devices.

\subsection{Related Works}
In \cite{nguyen2020spectral}, authors studied the IBFD-CF in the context of the centralized CF-mMIMO systems, where a small amount of IAI is mitigated in digital domain by the coordinated precoding and successive interference cancellation operated at APs' receiver. With this method, each AP has to reduce its transmit power to decrease the effect of IAI on the neighboring APs, which can result in performance degradation. On top of \cite{nguyen2020spectral}, the authors of \cite{xia2021joint,wang2019performance} proposed a so-called network-assisted full duplex (NAFD) mode for the centralized CF networks, where digital-domain IAI and IMI are mitigated by the central processing unit (CPU) and user scheduling, respectively, while the analog-domain CLI is not fully considered. In this case, the digital-domain IAI mitigation relies on both the relatively precise estimation of IAI channels and the undistorted fronthaul transmissions, which significantly increase system overhead. Possibly due to the above-mentioned challenges, to the best of our knowledge, there is no research article in the open literature having considered the full-duplex (FD)-style distributed CF systems, as the CLI problem in distributed CF systems is much severer than that in the centralized CF systems. 

To fill the research gap, in \cite{li2022Spectral}, we proposed a distributed CF-mMIMO system driven by the multicarrier-division duplex (MDD), which enables FD operation in the same time slot and the same frequency band but on different subcarriers. In our MDD-assisted CF (MDD-CF) systems, the subcarriers of one band are divided into two mutually exclusive subsets, namely DL and UL subcarrier subsets, to support DL and UL transmissions, respectively.  With the aid of these arrangements, despite the MDD-CF system suffers from the problems of SI and CLI in analog domain similarly as the IBFD-CF system, it can be free from the digital-domain SI and CLI. This can be readily achieved by the fast Fourier transform (FFT) operation at receiver, as the SI/CLI and desired signals in the MDD-CF system are transmitted on orthogonal subcarriers \cite{li2021multicarrier}. Our studies in \cite{li2022Spectral} demonstrated that the MDD-CF scheme can significantly enhance the system performance, when compared with the IBFD-CF and TDD-CF systems.

%Therefore, when leveraging the existing approaches in analog domain, such as power allocation (PA) and antenna polarization \cite{kim2020dynamic,kolodziej2019band}, and the FFT operation in digital domain, the CLI and SI can be efficiently mitigated at individual AP and MS node, enabling the FD-style distributed CF systems practical. 

However, the PA problem in the distributed MDD-CF system is typically non-convex and computationally challenging due to the large size of the optimization variables involved. In literature, there are various optimization algorithms proposed for resource allocation, including the Dinkelbach's transform \cite{you2021energy}, quadratic transform (QT) \cite{shen2018fractional,amudala2019spectral,li2022Spectral}, dual decomposition \cite{fang2019optimal} and the successive convex approximation (SCA) \cite{li2017optimal,xu2016power}. These approaches can in general converge to the local optimum within several iterations, but their computational complexity grows exponentially with the increase of network size. Moreover, once network changes, these algorithms have to rerun for obtaining the updated solutions, which introduces extra computational cost and latency. As the result, these traditional methods can hardly be operated in a timely manner for the PA in the CF-mMIMO systems. Needless to say, the situation is more complicated in the FD-style scenarios.

Recently, machine learning (ML) has shown its potential for solving the PA problem in wireless communications. For instance, in \cite{lee2018deep}, a convolutional neural network (CNN)-based PA algorithm was proposed and shown to be superior to the conventional weighted minimum mean square error (WMMSE) method in terms of the achievable SE and computation time. In \cite{liang2019towards}, the authors proposed an ensemble PA network (ePCNet) consisting of multiple independent and fully-connected multi-layer neural networks, which was demonstrated to outperform the WMMSE and greedy-relied PA. 
%In \cite{van2020power}, a CNN-based model combined with a residual structure was proposed in mMIMO systems, and is capable of allocating pilot and data power for a varying number of users. 
Furthermore, in \cite{luo2022downlink} and \cite{zhao2021dynamic}, the deep reinforcement learning-based PA was leveraged to solve respectively the max-min and SE maximization problems in the CF-mMIMO systems. In \cite{bashar2020deep}, the authors resorted to the deep CNN for maximizing SE in the CF-mMIMO systems, and demonstrated that the learning-based method can outperform the well-known use-and-then-forget-based PA method. In \cite{zaher2021learning}, a clustered DNN model exploiting the large-scale fading information was introduced to implement PA in the CF-mMIMO systems, showing that its performance is comparable to that of the WMMSE-assisted alternating direction method. 

The aforementioned learning-based PA methods have been shown to prevail over the traditional methods. However, none of them have considered to exploit the structures of wireless networks in the optimization process. Hence, they can not be generalized to the unseen scenarios, such as the networks with varying sizes or AP/MS densities. To this end, the graph neural network (GNN) reaping the advantages of scalability, generalization and parallel execution has attracted significant research interests in wireless communications. Specifically, in \cite{shen2020graph}, a wireless channel graph convolution network (WCGCN) was proposed for solving the PA and bearmforming problem in device-to-device (D2D) communication systems. It was shown that the WCGCN trained on the basis of small size systems can be generalized to use for optimization in the large systems with higher density of MSs and larger cell size. Additionally, the authors in \cite{chowdhury2021unfolding} and \cite{eisen2020optimal} studied the GNN-assisted PA in the ad hoc networks. However, the above-mentioned GNN-related models for ad hoc or D2D networks are all homogeneous, where only one type of nodes exist. Hence, they are not suitable for the operation in the more complicated scenarios like the MDD-CF networks, where APs and MSs are different types of nodes having different node features. Furthermore, in the MDD-CF networks, each AP/MS node can be connected with different types of nodes either via the communication path or via the interference path.

Against the background, in this paper, we propose a CF-heterogeneous graph neural network (HGNN), to solve the PA problem in the distributed MDD-CF systems. To the best of our knowledge, this is the first HGNN-based network proposed for handling the PA problem in the FD-style CF-mMIMO systems.

\subsection{Contributions}
The major contributions are summarized as follows:
\begin{itemize}

%\item We integrate the MDD mode with the CF systems to enable the FD-style operation in the distributed multicarrier mMIMO CF systems, where each AP independently implements DL/UL beamforming in ZF principle. The PA problem for maximizing SE is formulated under the constraints of QoS requirements, where the effect of residual SI and CLI is practically modeled.

\item The PA problem for maximizing SE in the distributed MDD-CF systems is formulated under the constraints of QoS requirements, where the effect of residual SI and CLI are practically modeled. In order to solve the optimization problem, we propose a quadratic transform and successive convex approximation (QT-SCA) optimization algorithm. In this QT-SCA algorithm, QT is used to first transform the optimization function, which belongs to the multiple-ratio concave-convex fractional programming (MRCCFP) problem, into an iterative convex problem. Then, SCA is employed to substitute the non-convex constraints imposed by QoS requirements with their approximated convex forms. Furthermore, the feasible initialization, convergence behavior and computational complexity of the QT-SCA algorithm are analyzed. 

\item To achieve the scalability and make the optimization general, we propose the CF-HGNN to solve the PA problem in the distributed MDD-CF systems. In our CF-HGNN, an adaptive node embedding layer and an adaptive output layer are implemented so that the CF-HGNN can adapt to the MDD-CF networks with various numbers of APs, MSs and subcarriers. Two types of meta-paths are defined, namely the data transmission path and interfering path, to connect the involved AP/MS nodes. Furthermore, to learn the importance factors of the information received at the AP/MS nodes via two meta-paths during message passing, the meta-path attention mechanism is implemented with the CF-HGNN.

\item We conduct a range of experiments to evaluate the effectiveness of the CF-HGNN for the PA problem in the distributed MDD-CF systems, where the CF-HGNN is trained in an unsupervised way using unlabeled data. Simulation results show that, without exploiting any prior information, the CF-HGNN is capable of learning nearly the same PA strategy as the QT-SCA algorithm, and it can significantly outperform the traditional greedy unfair method. Our studies also demonstrate that the CF-HGNN has the advantage over the QT-SCA in terms of computational complexity and operation time. Furthermore, the CF-HGNN exhibits the scalability to the MDD-CF networks with various numbers of nodes and subcarriers.  Specifically, assisted by user clustering, the CF-HGNN exhibits the scalability to the large-scale MDD-CF networks, which cover large area, have a big number of subcarriers and simultaneously support a large number of nodes.

\end{itemize}

The rest of the paper is organized as follows. In Section II, we describe the distributed MDD-CF systems, and formulate the optimization problem for PA. Then, the QT-SCA algorithm for PA is analyzed. In Section III, the CF-HGNN for PA in the distributed MDD-CF systems is detailed. Simulation results are presented and analyzed in Section IV. Finally, conclusions are summarized in Section V. 

%%% training 方式， 在cpu training，不需要信道信息，只需要等效信道增益

%%% 有了user group, 可以扩展到更大的网络 

\section{System Model}
\begin{figure}
\centering
\includegraphics[width=0.5\linewidth]{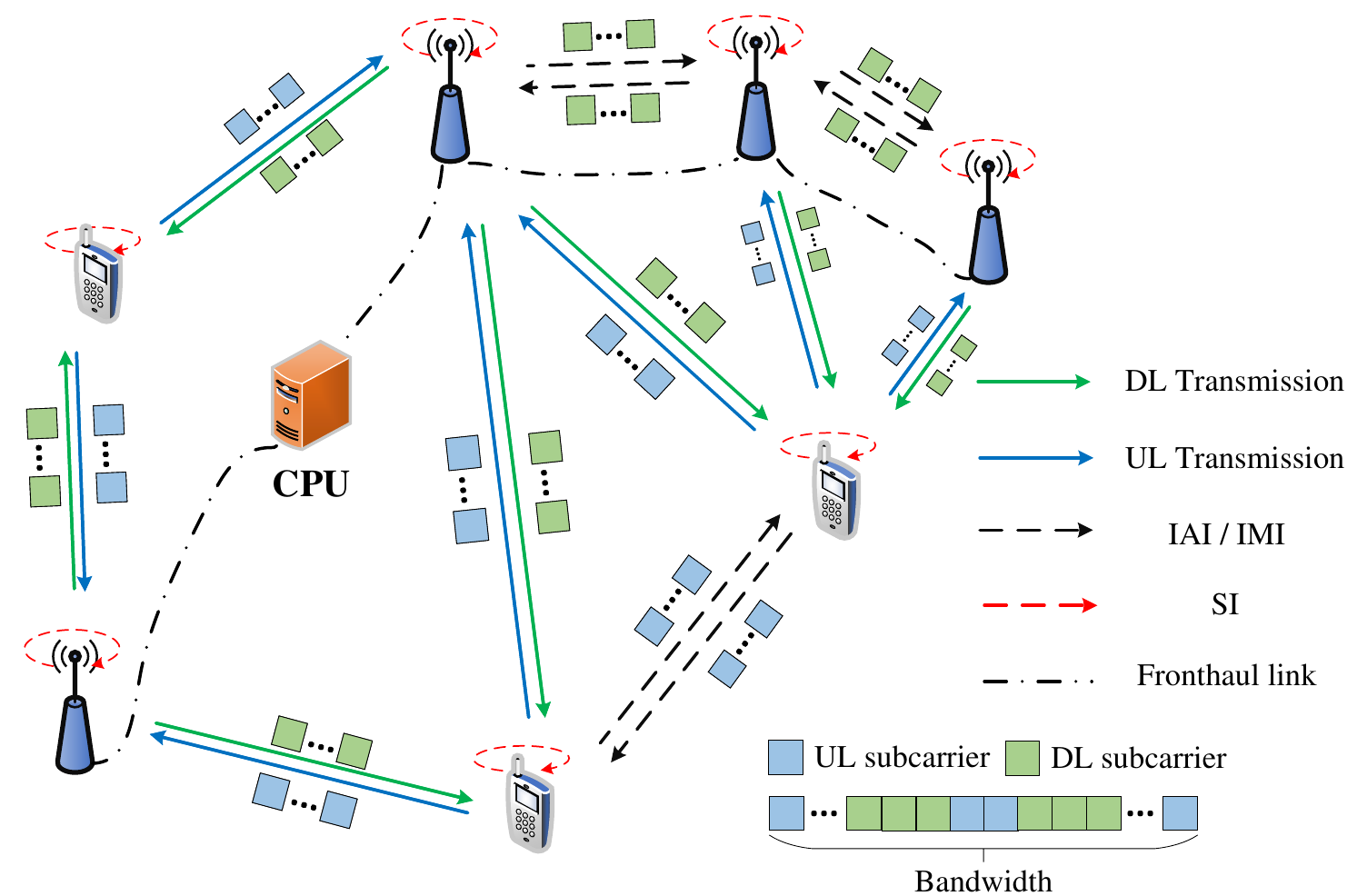}
\caption{Illustration of MDD-CF networks.}
\label{figure-MDDGNN-CF}
\end{figure}

An MDD-CF system as shown in Fig. \ref{figure-MDDGNN-CF} is studied. We assume the set $\mathcal{D}=\left\{1,...,d,...,D\right\}$ of single-antenna MSs and the set $\mathcal{L}=\left\{1,...,l,...,L\right\}$ of APs of each with $N$ antennas. The MSs and APs are operated in the MDD mode that relies on the mutually orthogonal subcarrier sets \cite{li2020self}, denoted as $\mathcal{M}=\left\{1,...,m,...,M\right\}$ with $\left|\mathcal{M}\right|=M$ and $\mathcal{\bar{M}}=\left\{1,...,\bar{m},...,\bar{M}\right\}$ with $\left|\mathcal{\bar{M}}\right|=\bar{M}$, for supporting  DL and UL data transmissions, respectively. The total number of subcarriers is $M_{\text{sum}} =M+\bar{M}$. Furthermore, we assume that the CF system is operated in the distributed manner, where CPU offloads the majority of tasks to APs to relieve its computation burden, and only sends the coded data to APs for DL transmissions or integrates the received UL data from APs via fronthaul links without requiring any channel knowledge.

\subsection{Channel Model}
%As in our proposed CF network, both APs and MSs transmit/receive data in MDD mode, the case of interference is much more complicated than TDD systems. 
For the convenience of notation, we denote the SI channels at the $l$-th AP and $d$-th MS by $\pmb{H}_{ll} \in \mathbb{C}^{N \times N}$ and $h_{dd}$, respectively. These two SI channels are modeled as
\begin{equation}\label{eq:MDDCF:SICh}
\left(\pmb{H}_{ll}\right)_{i,j}=\sqrt{\xi_l^{\text{SI}}} \ \alpha_s  , \ h_{dd} = \sqrt{\xi_d^{\text{SI}}} \ \alpha_s,
\end{equation}
where $\alpha_s\sim \mathcal{CN}(0,1)$ is the small-scale fading, while $\xi_l^{\text{SI}}\in(0,1]$ and $\xi_d^{\text{SI}}\in(0,1]$ denote the levels of the residual SI at AP and MS receivers, respectively.

Furthermore, we denote the time-domain channel impulse responses (CIRs) of the communication channels between the $d$-th MS and the $n$-th antenna at the $l$-th AP, the IAI channels between the $n$-th antenna at the $l$-th AP and the $n^{\prime}$-th antenna at the $l^{\prime}$-th AP, and the IMI channels between the $d$-th MS and $d^{\prime}$-th MS by $\pmb{g}_{ld}^n \in \mathbb{C}^{U_{ld} \times 1}$, $\pmb{g}_{ll^{\prime}}^ {nn^{\prime}}\in \mathbb{C}^{U_{ll^{\prime}} \times 1}$ and $\pmb{g}_{dd^{\prime}} \in \mathbb{C}^{U_{dd^{\prime}} \times 1}$, respectively, where $U\in\left\{U_{ld}, U_{ll^{\prime}}, U_{dd^{\prime}}\right\}$ is the number of taps of multipath channels. Specifically, the $u$-th tap of these channels is generally modeled as $\left(\pmb{g}\right)_u=\sqrt{\beta/U} \ \alpha_s$ with $\pmb{g} \in \left\{\pmb{g}_{ld}^n, \pmb{g}_{ll^{\prime}}^ {nn^{\prime}}, \pmb{g}_{dd^{\prime}} \right\}$, where $\beta \in\left\{\beta_{ld}, \beta_{ll^{\prime}}, \beta_{dd^{\prime}}\right\}$ accounts for the large-scale fading of path loss and shadowing. Different channel taps are assumed to be independent. Given the time-domain CIRs, the frequency-domain subcarrier channels can be obtained as $\pmb{h}=\pmb{F}\pmb{\Psi}\pmb{g}$, where $\pmb{h} \in \left\{\pmb{h}_{ld}^n, \pmb{h}_{ll^{\prime}}^ {nn^{\prime}}, \pmb{h}_{dd^{\prime}} \right\}$, $\pmb{F} \in \mathbb{C}^{M_{\text{sum}}\times M_{\text{sum}}}$ is the FFT matrix, and $\pmb{\varPsi}\in \mathbb{C}^{{M_{\text{sum}}\times U}}$ is constructed by the first $U$ columns of $\pmb{I}_{M_\text{sum}}$. Moreover, the single DL/UL subcarrier channel can be expressed as $h[m]=\pmb{\phi}_{\text{DL}}^T\pmb{h}$ and $h[\bar{m}]=\pmb{\phi}_{\text{UL}}^T\pmb{h}$, respectively, where $\pmb{\phi}_{\text{DL}}=\pmb{I}_{M_\text{sum}}^{(:,m)}$ and $\pmb{\phi}_{\text{UL}}=\pmb{I}_{M_\text{sum}}^{(:,\bar{m})}$ are mapping vectors. 
%Note that here $h[m]$ or $h[\bar{m}]$ denotes the point-to-point subcarrier channel, which will be further integrated into the vector and matrix for the AP-AP channel (i.e., $\pmb{H}_{ll^{\prime}}[m \  \text{or} \ \bar{m}] \in \mathbb{C}^{N \times N}$ ), AP-MS channel (i.e., $\pmb{h}_{ld}[m] \in \mathbb{C}^{N\times 1}$ and $\pmb{h}_{ld}[\bar{m}] \in \mathbb{C}^{N\times 1}$), respectively.

\subsection{Downlink Transmission}
The data transmitted on the $m$-th DL subcarrier for the $d$-th MS is denoted by $x_d[m]$, which satisfies $\mathbb{E}\left\{\left|x_d[m]\right|^2\right\}=1$. The transmitted signal on the $m$-th DL subcarrier by the $l$-th AP is given by
\begin{equation}\label{eq:MDD_CF:slm}
\pmb{s}_l[m]=\sum_{d \in \mathcal{D}} \sqrt{p_{ldm}}\pmb{f}_{ld}[m] x_d[m],
\end{equation} 
where $\pmb{f}_{ld}[m] \in \mathbb{C}^{N \times 1}$ denotes the precoding vector with $\left\|\pmb{f}_{ld}[m]\right\|_2^2=1$, and $p_{ldm}$ is the power allocated to the $d$-th MS on the $m$-th subcarrier. The total power budget at the $l$-th AP is expressed as $P_l$, satisfying $\sum_{m \in \mathcal{M}}\sum_{d \in \mathcal{D}} p_{ldm}\leq P_l$.

The signal received from the $m$-th DL subcarrier by the $d$-th MS can be expressed as
\begin{align}\label{eq:MDD_CF:ydm}
y_d[m] = \sum_{l \in \mathcal{L}} \pmb{h}_{ld}^H[m] \pmb{s}_l[m] + z^{\text{SI}}_d + z^{\text{IMI}}_d  + n_d,
\end{align}
where $n_d \sim \mathcal{CN}(0,\sigma^2)$ is the additive white Gaussian noise (AWGN). According to \cite{day2012full2,ng2016power}, the residual interference in the digital domain arising from SI and IMI \big(i.e., $z^{\text{SI}}_d$ and $z^{\text{IMI}}_d$ in \eqref{eq:MDD_CF:ydm}\big) can be modeled as Gaussian noise. Specifically, we express $z^{\text{SI}}_d \sim \mathcal{CN}(0,\mathbb{E}\left[\bar{z}^{\text{SI}}_d\left(\bar{z}^{\text{SI}}_d\right)^\ast\right])$ with $\bar{z}^{\text{SI}}_d=h_{dd}\sum_{\bar{m} \in \bar{\mathcal{M}}}\sqrt{p_{d\bar{m}}}x_d[\bar{m}]$, where $x_d[\bar{m}]$ denotes the data transmitted on the $\bar{m}$-th UL subcarrier by the $d$-th MS, $p_{d\bar{m}}$ denotes the transmitted power. $z^{\text{IMI}}_d \sim \mathcal{CN}(0,\xi_d^{\text{IMI}}\mathbb{E}\left[\bar{z}^{\text{IMI}}_d\left(\bar{z}^{\text{IMI}}_d\right)^\ast\right])$ with $\bar{z}^{\text{IMI}}_d=\sum_{d^{\prime} \in \mathcal{D}\backslash \left\{d\right\}}\sum_{\bar{m} \in \bar{\mathcal{M}}}\sqrt{p_{d^{\prime}\bar{m}}}h_{dd^{\prime}}[\bar{m}]x_{d^{\prime}}[\bar{m}]$, where $\xi_d^{\text{IMI}}$ denotes the residual IMI level at MS $d$.

Based on \eqref{eq:MDD_CF:ydm}, the received SINR on the $m$-th DL subcarrier at the $d$-th MS is given by
\begin{equation}\label{eq:MDD_CF:dlSINR}
\text{SINR}_{d,m} = \frac{\left|\sum_{l \in \mathcal{L}}\sqrt{p_{ldm}}\pmb{h}_{ld}^H[m]\pmb{f}_{ld}[m]\right|^2}{\text{MUI}_{d,m}+ \text{var}\left\{z^{\text{SI}}_d\right\} + \text{var}\left\{z^{\text{IMI}}_d\right\}+\sigma^2} \ ,
\end{equation} 
where $\text{MUI}_{d,m}=\sum_{l \in \mathcal{L}}\sum_{d^{\prime} \in \mathcal{D}\backslash \left\{d\right\}}p_{ld^{\prime}m}|\pmb{h}_{ld}^H[m]\pmb{f}_{ld^{\prime}}[m]|^2$ is the component of multiuser interference (MUI).

\subsection{Uplink Transmission}
The received UL signal by the $l$-th AP from the $\bar{m}$-th subcarrier can be expressed as
\begin{equation}\label{eq:MDD_CF:ylbarm}
\pmb{y}_l[\bar{m}] = \sum_{d\in\mathcal{D}}\sqrt{p_{d\bar{m}}}\pmb{h}_{ld}[\bar{m}]x_d[\bar{m}]+ \pmb{z}^{\text{SI}}_l + \pmb{z}^{\text{IAI}}_l + \pmb{n}_l,
\end{equation}
where $p_{d\bar{m}}$ denotes the power allocated by MS $d$ to the $\bar{m}$-th UL subcarrier, which satisfies $\sum_{\bar{m} \in \bar{\mathcal{M}}}p_{d\bar{m}}\leq P_d$. Similar to the received signals at MSs, the residual interferences due to the SI and IAI are modeled as Gaussian noise, where $\pmb{z}^{\text{SI}}_l \sim \mathcal{CN}(0,\text{diag}(\mathbb{E}[\bar{\pmb{z}}^{\text{SI}}_l(\bar{\pmb{z}}^{\text{SI}}_l)^H]))$ with $\bar{\pmb{z}}^{\text{SI}}_l=\pmb{H}_{ll}\sum_{m \in \mathcal{M}}\pmb{s}_l[m]$, and $\pmb{z}^{\text{IAI}}_l \sim \mathcal{CN}(0,\xi_l^{\text{IAI}}\text{diag}(\mathbb{E}[\bar{\pmb{z}}^{\text{IAI}}_l(\bar{\pmb{z}}^{\text{IAI}}_l)^H]))$ with $\bar{\pmb{z}}^{\text{IAI}}_l=\sum_{l^{\prime} \in \mathcal{L}\backslash \left\{l\right\}}\sum_{m \in \mathcal{M}}\pmb{H}_{ll^{\prime}}[m]\pmb{s}_{l^{\prime}}[m]$, where $\xi_l^{\text{IAI}}$ denotes the residual IAI level at the AP side.

Due to the distributed processing nature in our proposed system, each AP firstly processes the signal received from MSs using the local combining vectors, expressed as $\tilde{y}_l[\bar{m}]=\pmb{w}_{ld}^H[\bar{m}]\pmb{y}_l[\bar{m}]$, where $\pmb{w}_{ld}[\bar{m}]$ denotes the local combining vector of AP $l$ for detecting MS $d$. Then, the locally estimated data by all APs are forwarded to the CPU for final processing, which can be expressed as $y_{\text{cpu}}[\bar{m}]=\sum_{l \in \mathcal{L}}\tilde{y}_l[\bar{m}]$. It can be shown that the SINR at the CPU for detecting the data transmitted on the $\bar{m}$-th UL subcarrier of MS $d$ can be expressed as
%\begin{figure*}[b]
\begin{equation}\label{eq:MDD_CF:ulSINR}
\text{SINR}_{d,\bar{m}}=\frac{p_{d\bar{m}}|\tilde{\pmb{w}}_d[\bar{m}]\tilde{\pmb{h}}_d[\bar{m}]|^2}{\text{MUI}_{d,\bar{m}}+ \text{SI}_{d,\bar{m}}+\text{IAI}_{d,\bar{m}}+\sigma^2\left\|\tilde{\pmb{w}}_d[\bar{m}]\right\|^2} \ ,
\end{equation}  
%\end{figure*}
where $\tilde{\pmb{w}}_d[\bar{m}]=[\pmb{w}_{1d}^H[\bar{m}],...,\pmb{w}_{Ld}^H[\bar{m}]]$, $\tilde{\pmb{h}}_d[\bar{m}]=[\pmb{h}_{1d}^H[\bar{m}],...,\pmb{h}_{Ld}^H[\bar{m}]]^H$, \\ $\text{MUI}_{d,\bar{m}}=\sum_{d^{\prime} \in \mathcal{D}\backslash \left\{d\right\}}p_{d^{\prime}\bar{m}}|\tilde{\pmb{w}}_d[\bar{m}]\tilde{\pmb{h}}_{d^{\prime}}[\bar{m}]|^2$, $\text{SI}_{d,\bar{m}}=\sum_{l \in \mathcal{L}}\mathbb{E}[\left\|\pmb{w}_{ld}^H[\bar{m}]\pmb{z}^{\text{SI}}_l\right\|^2]$ and  \\ $\text{IAI}_{d,\bar{m}}=\sum_{l \in \mathcal{L}}\mathbb{E}[\left\|\pmb{w}_{ld}^H[\bar{m}]\pmb{z}^{\text{IAI}}_l\right\|^2]$.
%\begin{remark}
%As mentioned in Section \ref{section:MDDCF:intro}, in IBFD-, DTDD- and NAFD-CF multicarrier systems, the IMI and IAI belong to the co-subchannel interference, and as APs employed with small number of antennas are usually densely distributed, these interferences can not be simply assumed as Gaussian noise in digital domain. The reason is that if APs are close to each other, the large-scale fading is insufficient to suppress IAI under the level of background noise. Consequently, the huge power gap between IAI and UL signal introduces large quantization noise after ADC operation at receiver, and hence desired signal is hardly detected. This analysis also holds for IMI case. 
%\end{remark}

\subsection{Beamforming Design}\label{sub:MDD_GNN:Beam}
In this paper, the ZF beamforming strategy is employed for both transmitting and receiving at APs. It is well-known that the MMSE beamforming slightly outperforms the ZF beamforming in mMIMO, when perfect CSI is available and especially when SINR is low \cite{bjornson2017massive}. However, considering the multiuser interference suppression, computation complexity as well as the concise formulation, ZF is considered in the following analysis. %Nevertheless, the MMSE can be equally applied in our proposed network. 
%As for maximum-ratio (MR) beamforming, since the distributed operation is assumed, APs employed with few antennas work individually, and then MR has the limited performance compared with ZF and MMSE \cite{demir2021foundations}.} 

According to the ZF principle \cite{jiang2011performance}, the precoder/combiner at the $l$-th AP, expressed as $\pmb{F}^{\text{ZF}}_l[m]=[\pmb{f}_{l1}^{\text{ZF}}[m],...,\pmb{f}_{lD}^{\text{ZF}}[m]]$ and $\pmb{W}^{\text{ZF}}_l[\bar{m}]=[\pmb{w}_{l1}^{\text{ZF}}[\bar{m}],...,\pmb{w}_{lD}^{\text{ZF}}[\bar{m}]]$, can be derived as $\pmb{F}^{\text{ZF}}_l[m]=\pmb{H}_{l}^H[m](\pmb{H}_{l}[m]\pmb{H}_{l}^H[m])^{-1}$ and $\pmb{W}^{\text{ZF}}_l[\bar{m}]=\pmb{H}_{l}[\bar{m}](\pmb{H}_{l}^H[\bar{m}]\pmb{H}_{l}[\bar{m}])^{-1}$, respectively, where $\pmb{H}_{l}[m]=[\pmb{h}_{l1}[m],...,\pmb{h}_{lD}[m]]^H$ and  $\pmb{H}_{l}[\bar{m}]=[\pmb{h}_{l1}[\bar{m}],...,\pmb{h}_{lD}[\bar{m}]]$. Note that, in order to ensure that the MUI is fully suppressed, the implementation of ZF beamforming should adhere to the constraint of $N\geq D$ \footnote{We will discuss the case of $D\gg N$ in Section \ref{sec:MDD_GNN:case2} along with the user-centric clustering algorithm.}.
%\footnote{For the sake of convenience, we assume that each AP is employed with sufficient antennas so as to suppress the interference that itself generates. Although an AP is expected to be equipped with a small number of antennas in CF networks, our assumption is still practical, as each AP can be treated as a secondary central unit controlling $N$ single-antenna AP operated in a centralized mode through fronthaul connections.} 
In this case, the MUI terms in \eqref{eq:MDD_CF:dlSINR} and \eqref{eq:MDD_CF:ulSINR} are equal to zero. Therefore, the $\text{SINR}_{d,m}$ and $\text{SINR}_{d,\bar{m}}$ can be rewritten as
\begin{align}\label{eq:MDDGNN:SINRSim}
\text{SINR}_{d,m}&=\frac{|\sum_{l \in \mathcal{L}}\sqrt{p_{ldm}}\omega_{ldm}|^2}{\xi_d^{\text{SI}}\Theta_{\text{DL}}+\sigma^2}, \nonumber\\
\text{SINR}_{d,\bar{m}}&=\frac{p_{d\bar{m}}L^2}{\sum_{l \in \mathcal{L}}\upsilon_{ld\bar{m}}(\xi_l^{\text{SI}}\Theta_{\text{UL}}+\sigma^2)} ,
\end{align} 
where $\omega_{ldm}=1/{\|\pmb{f}_{ld}^{\text{ZF}}[m]\|_2}$, $\upsilon_{ld\bar{m}}=\|\pmb{w}_{ld}^{\text{ZF}}[\bar{m}]\|_2^2$, \\ $\Theta_{\text{DL}}=\sum_{\bar{m} \in \bar{\mathcal{M}}}p_{d\bar{m}} +\sum_{d^{\prime} \in \mathcal{D}\backslash \left\{d\right\}}\sum_{\bar{m} \in \bar{\mathcal{M}}}(\xi_d^{\text{IMI}}\beta_{dd^{\prime}}p_{d^{\prime}\bar{m}}/{\xi_d^{\text{SI}}}{M_{\text{sum}}})$ and  $\Theta_{\text{UL}}=\sum_{m \in \mathcal{M}}\sum_{d \in \mathcal{D}}  p_{ldm}+\sum_{l^{\prime} \in \mathcal{L}\backslash \left\{l\right\}}\sum_{m \in \mathcal{M}}\sum_{d \in \mathcal{D}}(\xi_l^{\text{IAI}}\beta_{ll^{\prime}}p_{l^{\prime}dm}/{\xi_l^{\text{SI}}}{M_{\text{sum}}})$. 
%For the details of simplification, please refer to Appendix \ref{Appen:MDD_CF_SINRSim}.

\subsection{Problem Formulation}
Given the $\text{SINR}_{d,m}$ and $\text{SINR}_{d,\bar{m}}$ as shown in \eqref{eq:MDDGNN:SINRSim}, the optimization problem can be formulated as
\begin{subequations}
\label{eq:MDDGNN:SE_formulation}
\begin{align}
&\max_{p_{ldm}, p_{d\bar{m}}} \Lambda_{\text{SE}} \\ 
\text{s.t.} \ \ &~~\sum_{m \in \mathcal{M}}\sum_{d \in \mathcal{D}} p_{ldm}\leq P_l, \ \forall l \in \mathcal{L}, \\
&~~\sum_{\bar{m} \in \bar{\mathcal{M}}}p_{d\bar{m}}\leq P_d, \ \forall d \in \mathcal{D}, \\
&~~\sum_{m \in \mathcal{M}}\ln(1+\text{SINR}_{d,m}) \geq \chi_{\text{DL}}, \ \forall d \in \mathcal{D}, \\
&~~\sum_{\bar{m} \in \mathcal{\bar{M}}}\ln(1+\text{SINR}_{d,\bar{m}}) \geq \chi_{\text{UL}}, \ \forall d \in \mathcal{D},
\end{align}
\end{subequations}
where $\Lambda_{\text{SE}}=\frac{1}{M_{\text{sum}}}\sum_{d \in \mathcal{D}} \big(\sum_{m \in \mathcal{M}}\ln(1+\text{SINR}_{d,m}) +\sum_{\bar{m} \in \mathcal{\bar{M}}}\ln(1+\text{SINR}_{d,\bar{m}})\big)$, $\chi_{\text{DL}}$ and $\chi_{\text{UL}}$ denote the DL and UL QoS requirements, respectively. 

It can be observed that (\ref{eq:MDDGNN:SE_formulation}) is a NP-hard nonconvex problem with the nonconvex constraints as shown in (\ref{eq:MDDGNN:SE_formulation}d) and (\ref{eq:MDDGNN:SE_formulation}e). Generally, to solve this kind of complicated problems, we have to transform the main optimization function as well as the constraints into the approximated convex ones, and then solve them in an iterative way. With this consideration, in this paper, we propose a QT-SCA method, which can be found in detail in Appendix \ref{app:MDDGNN:QTSCA}. Unfortunately, the complexity of QT-SCA increases significantly as the size of network becomes large, especially when the multicarrier FD-like systems, such as our proposed MDD-CF system, are considered. To this end, we resort to the GNN to solve this optimization problem, which will be demonstrated to exhibit low complexity and high efficiency.

\section{Graph Learning in MDD-CF networks}
In order to leverage the structural information of MDD-CF network to solve the complicated problem, as that formulated in \eqref{eq:MDDGNN:SE_formulation}, the heterogeneous graph learning based PA scheme is introduced. We aim to learn a scalable and transferable HGNN to efficiently distribute both APs' and MSs' transmit power to maximize the SE of the distributed MDD-CF systems.
\subsection{Definition of Heterogeneous Graph}
A heterogeneous graph can be represented as $G=(\mathcal{V},\mathcal{E})$, where $\mathcal{V}$ is the set of nodes, and $\mathcal{E}$ is the set of edges. The heterogeneous graph has a node type mapping function $\phi: \mathcal{V}\longrightarrow \mathcal{Q}$ and an edge type mapping function $\psi: \mathcal{E}\longrightarrow \mathcal{P}$, where $\mathcal{Q}$ and $\mathcal{P}$ denote the sets of predefined node types and link types, $\left|\mathcal{Q}\right|+\left|\mathcal{P}\right|>2$ \cite{sun2013mining}. Explicitly, we can write $\mathcal{Q}=\{Q_1,Q_2,...\}$ and $\mathcal{P}=\{P_1,P_2,...\}$, where $Q_i$ and $P_j$ are the $i$-th node type and $j$-th link type. Let $\pmb{v}_i\in\mathbb{R}^{F_v\times 1}$ denote a node with $F_v$-dimensional features and $\pmb{e}_{i,j}\in\mathbb{R}^{F_e\times 1} $ denote an edge pointing from $\pmb{v}_j$ to $\pmb{v}_i$, which has $F_e$-dimensional features. Given the mapping function $\phi$ and $\psi$, each node belongs to a particular node type of $\phi(\pmb{v})\in \mathcal{Q}$ and each edge belongs to a specific relation of $\psi(\pmb{e})\in \mathcal{P}$. The neighborhood of a node $\pmb{v}_i$ is defined as $\mathcal{N}_i=\left\{\pmb{v}_j\in \mathcal{V}|\pmb{e}_{i,j}\in\mathcal{E}\right\}$. Furthermore, in a heterogeneous graph, two nodes may be connected via different semantic paths. For example, an AP and an MS can be connected via two paths, namely the AP-DL-MS and MS-UL-AP links. Therefore, we introduce the concept of meta-path as in \cite{wang2019heterogeneous}. In detail, a meta-path $\Phi$ is defined as a path in the form of $Q_1\stackrel{P_1}{\longrightarrow}Q_2\stackrel{P_2}{\longrightarrow}\cdots \stackrel{P_{n}}{\longrightarrow}Q_{n+1}$, which defines a composite relation $P=P_1\circ P_2 \circ \ldots P_n$ from node type $Q_1$ to node type $Q_{n+1}$, where $\circ$ denotes the composition operator on relations. Once a meta-path $\Phi$ is given, the specific neighbors $\mathcal{N}_i^\Phi$ of node $\pmb{v}_i$ can be obtained, which are given by the set of nodes connected with $\pmb{v}_i$ via meta-path $\Phi$.

\begin{figure}
\centering
\includegraphics[width=0.5\linewidth]{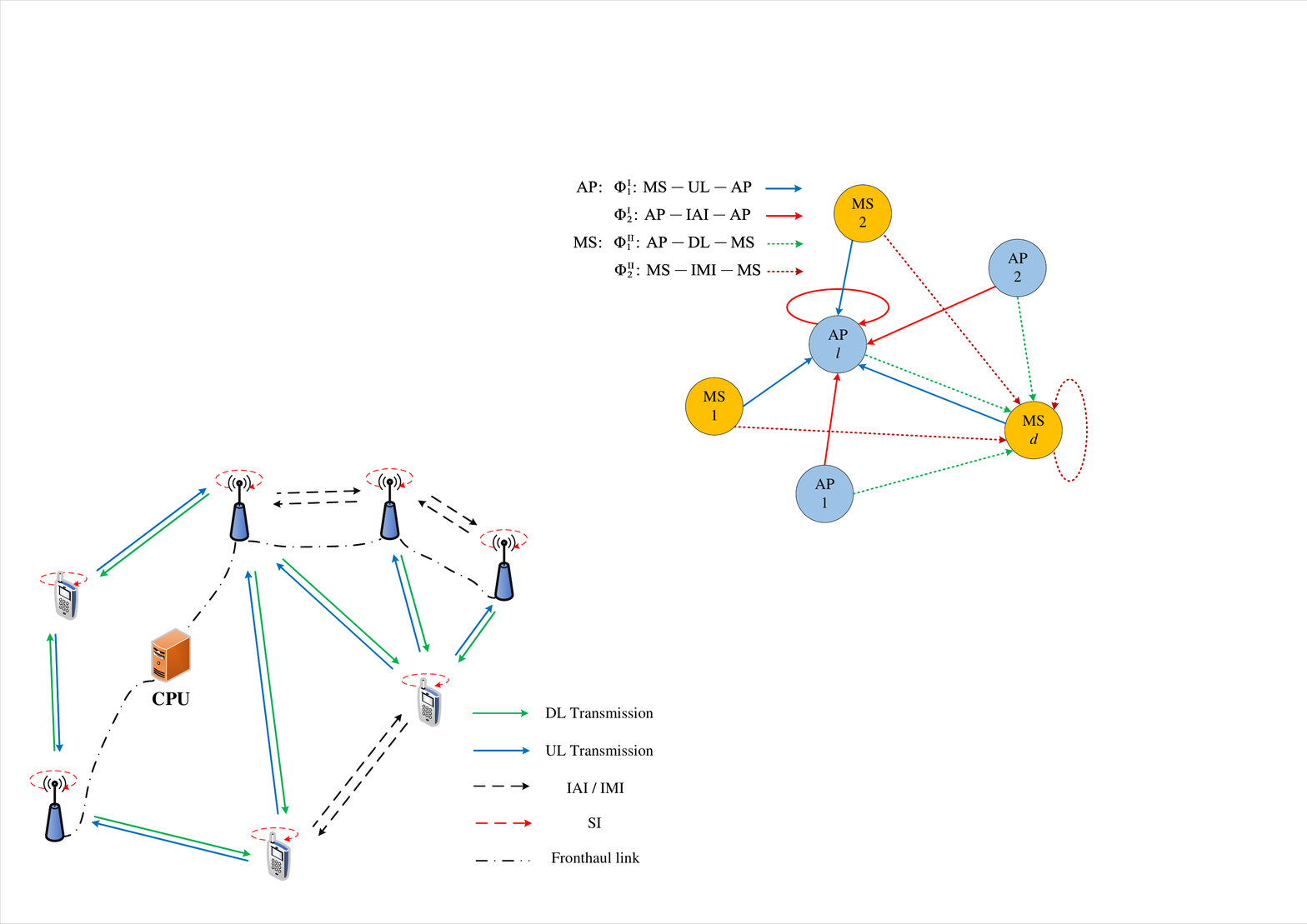}
\caption{An exemplified graph of CF network. }
\label{figure-MDDCGNN-model}
\end{figure}

\subsection{Heterogeneous Graph for MDD-CF Networks}
Intuitively, as shown in Fig. \ref{figure-MDDCGNN-model}, it is straightforward to model an MDD-CF network as a heterogeneous graph. It can be observed from the figure that there are two types of nodes, namely AP and MS, each of which is associated with two meta-paths. Specifically, for AP $l$, all MSs are connected with it via the meta-path $\Phi_1^{\text{I}}$ (MS-UL-AP), and the remaining APs are connected with it via the meta-path $\Phi_2^{\text{I}}$ (AP-IAI-AP). Note that, the SI caused by AP $l$ itself is classified into $\Phi_2^{\text{I}}$ by adding a self-loop. Similarly, the meta-paths $\Phi_1^{\text{II}}$ and $\Phi_2^{\text{II}}$ associated with MS $d$ are AP-DL-MS and MS-IMI-MS, respectively. 

In the heterogeneous graph of MDD-CF, the node feature vectors of AP $l$ and MS $d$ are defined as
\begin{align}\label{eq:MDDGNN:nodef}
\pmb{v}_l \in \mathbb{R}^{(DM + 3)\times1} &= \left[\pmb{\omega}_{l1}^T...\pmb{\omega}_{lM}^T, P_l, \xi_l^{\text{SI}}, \xi_l^{\text{IAI}}\right]^T, \nonumber \\
\pmb{v}_d \in \mathbb{R}^{(L\bar{M} + 3)\times1} &= \left[\pmb{\upsilon}_{1d}^T...\pmb{\upsilon}_{\bar{M}d}^T, P_d, \xi_d^{\text{SI}}, \xi_d^{\text{IMI}}\right]^T,
\end{align}
respectively, where $\pmb{\omega}_{lm} = \left[{\omega}_{l1m},...,{\omega}_{lDm}\right]^T$ is the $m$-th DL equivalent subchannel gains between AP $l$ and all the $D$ MSs, and $\pmb{\upsilon}_{\bar{m}d} = \left[{\upsilon}_{1d\bar{m}},...,{\upsilon}_{Ld\bar{m}}\right]^T$ is the $\bar{m}$-th UL equivalent subchannel gains between MS $d$ and all the $L$ APs, when the ZF beamforming as in equation \eqref{eq:MDDGNN:SINRSim} is applied. Moreover, to simplify the model for the sake of reducing complexity, the attribute of edges is assumed to be the Euclidean distance between any two nodes, expressed as $e_{i,j} = d_{i,j}, \forall i, j \in \mathcal{V}$. If $i=j, e_{i,j}=0$ denotes the edge feature of self-loop. 
\begin{remark}
In the distributed MDD-CF systems, each AP equipped with a baseband processor is able to independently implement DL/UL beamforming, and the CPU only needs to collect the processed signals from APs and then accomplish the final data detection. Hence, in order to fully exploit the APs' computational potentials while avoiding using long-stacked channel vectors as node features, we assume that, each AP first computes the equivalent DL/UL subchannel gains based on the estimated CSI. Then, the equivalent DL/UL subchannel gains are transmitted to the CPU as the AP/MS node features. All these are done during the offline graph training\footnote{In principle, apart from ZF, the beamforming methods, such as MMSE or matched filtering, may also be employed. If the number of antennas at individual AP and the numbers of MSs and subcarriers are small, it is possible to represent the AP and MS node features by their channel vectors, which have the size of $(2DNM+3)$ and $(2LN\bar{M}+3)$, respectively, when both the real and imaginary parts of the complex vectors are considered. Correspondingly, the beamforming vectors associated with PA can be directly learned at the final output layer, rather than choosing one of the known beamforming schemes to generate the network inputs. However, if the numbers of antennas at each AP, MSs and subcarriers are large, the huge dimensions of node features will make the model extremely hard to train.}. Note that in some references, such as \cite{cui2019spatial}, the authors used the geographic location information (GLI) as the input of the learning-based network to reduce the training overhead. This approach has the advantage that the low-dimensional coordinate values of AP/MS can be directly used as node features without requiring to computing beamformers. However, in MDD-based systems, power needs to be allocated among different subcarriers, while the GLI lacks not only the small-scale information of the involved communication channels but also the multi-antenna characteristics of MIMO systems. Explicitly, the GLI is unable to provide the required information and hence, is infeasible in MDD-based systems.
\end{remark} 

\subsection{Heterogeneous Graph Learning Assisted Power Allocation in MDD-CF Networks}
In this subsection, we formally present the CF-HGNN to solve the PA problem in MDD-CF networks. The architecture of CF-HGNN consists of four components: 1) adaptive node embedding; 2) meta-path based message passing; 3) meta-path based attention; 4) downstream PA learning. The overall CF-HGNN is type-specific and the parameters for processing AP and MS nodes are not shared.

\begin{figure}
\centering
\includegraphics[width=0.5\linewidth]{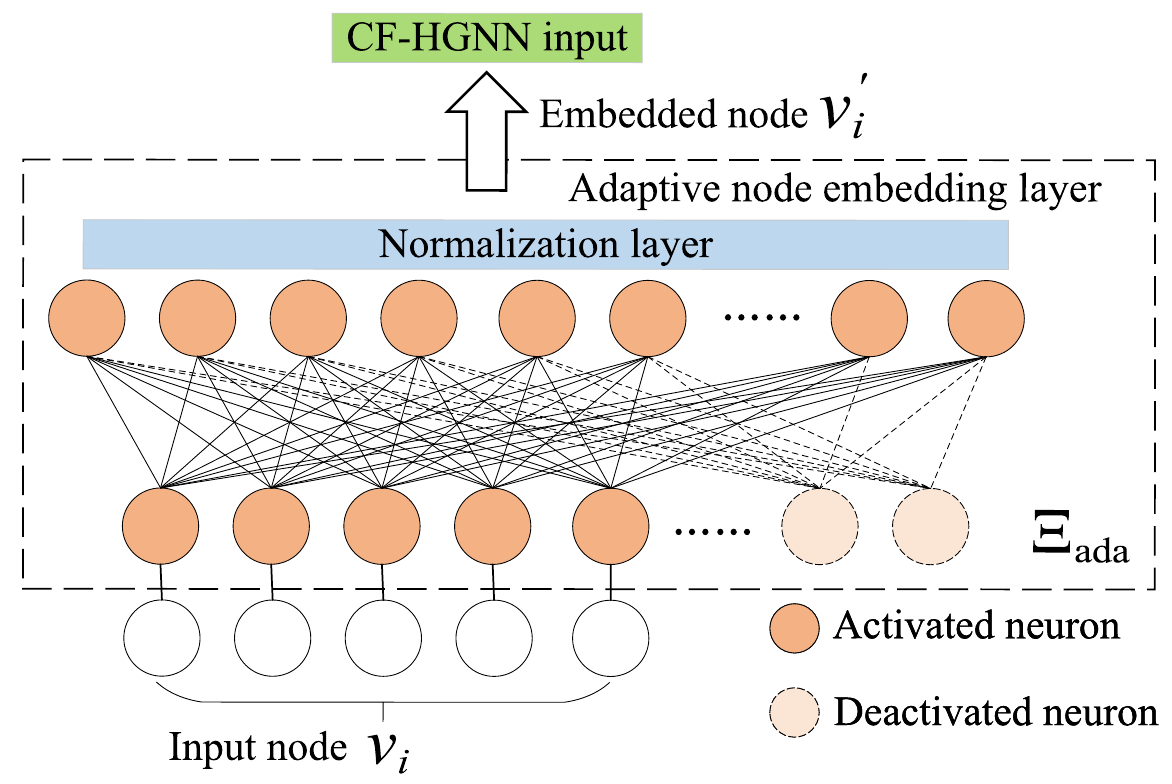}
\caption{Adaptive node embedding layer.}
\label{figure-MDDCGNN-node}
\end{figure}

\subsubsection{Adaptive Node Embedding}
In general, in order to guarantee the scalability of GNN, the size of node feature should be irrelevant to the number of nodes involved. As we can see in \eqref{eq:MDDGNN:nodef}, the feature dimensions of input nodes are related to $L$ and $D$, which vary with respect to the scales of CF networks. Hence, they are not feasible to exhibit the scalability and transferability. To tackle this problem, we propose an adaptive node embedding layer, which can handle graph nodes with varying input feature dimensions through adaptively activating or deactivating the neurons in the multi-layer perceptron (MLP), i.e., $\Xi_{\text{ada}}(\pmb{v}_l)$ and $\Xi_{\text{ada}}(\pmb{v}_d)$, as depicted in Fig. \ref{figure-MDDCGNN-node}. Specifically, before the normalization layer, AP and MS nodes are transformed by two embedding matrices, which are expressed as 
\begin{align}\label{eq:MDDGNN:embed}
\pmb{v}_l^{'} &= \pmb{W}_{\text{AP}}^{(:,1:DM+3)} \pmb{v}_l,  \nonumber \\
\pmb{v}_d^{'} &= \pmb{W}_{\text{MS}}^{(:,1:L\bar{M}+3)} \pmb{v}_d,
\end{align}
where $\pmb{W}_{\text{AP}} \in \mathbb{R}^{F_{\text{AP}}^{'}\times F_{\text{AP}} }$ and $\pmb{W}_{\text{MS}} \in \mathbb{R}^{F_{\text{MS}}^{'}\times F_{\text{MS}} }$ map different AP and MS input nodes into two feature domains with predefined sizes of $F_{\text{AP}}^{'}$ and $F_{\text{MS}}^{'}$. Note that, as the ZF beamforming is applied, the maximum input size of the embedding layer for AP nodes is subject to the number of antennas employed at individual AP, having the relationship of $F_{\text{AP}}= NM+3$. In the context of MS, we assume $F_{\text{MS}}= L^{'}\bar{M}+3$, where $L^{'}$ denotes the maximum number of APs that can be deployed in a CF network within a certain area. %Therefore, as long as the antenna configuration and AP density are constant, the CF-HGNN model can be applied in any CF network with varying number of MSs and APs.
\begin{remark}\label{remark:MDD_GNN:re2}
In Section \ref{sub:MDD_GNN:Beam}, we assumed that, in our proposed distributed MDD-CF system, each AP having $N\geq D$ antennas individually communicates with all the $D$ MSs using ZF beamforming. In this case, the above-mentioned embedding layer imposes strict limitation on the maximum number of MSs, which should not exceed the number of antennas configured at each AP. However, if MSs are densely distributed or each AP is only equipped a small number of antennas, the AP node features can no longer be set as \eqref{eq:MDDGNN:nodef}, as the ZF beamforming cannot be achieved in the case of $D>N$ \footnote{Although the MMSE beamforming can be applied in the case of $D>N$, the increasing number of MSs leads to large multiuser interference and high dimensions of AP node features.}. Moreover, in the above model, the number of APs is restricted to $L^{'}$, meaning that the CF-HGNN cannot be generalized to the CF networks with more than $L^{'}$ APs \footnote{In principle, $L^{'}$ can be predefined to a sufficiently large number so as to make CF-HGNN scalable to the network with densely distributed APs. However, in doing so, it will significantly increase the training overhead of the CF-HGNN.}. In these cases, to guarantee the scalability of CF-HGNN and reduce the computational complexity, one possible approach is to transform the dense graph into the sparse graph with the aid of user-centric clustering \cite{demir2021foundations}, where each AP only serves a certain number of MSs. For example, if ZF beamforming is used, each AP can serve up to $N$ MSs depending on the channel conditions. Given this constraint, the dimensions of the AP node features are only related to the number of antennas, while that of the MSs node features are only relied on a fraction of APs. Hence, our node feature definition of \eqref{eq:MDDGNN:nodef} is still applicable. In Section \ref{sec:MDD_GNN:case2}, we will present an example accounting for the user-centric clustering.  %Another possible method is changing the way of setting the node features. For instance, in the case of small number of antennas employed at each AP,  
%To stay focused, we only consider the case of $N \geq D$, and the user-centric clustering will be considered in future research. 

%Generally, in order to guarantee the scalability of GNN, the size of node feature should be irrelevant to the number of nodes. To this end, since we assume $N\geq D$ and $L^{'}\geq L$, with the aid of adaptive node embedding, the dimension of AP and MS nodes input are only associated with the number of antennas employed at each AP and maximum number of AP deployed at CF networks. This is legitimate if abundant antennas can be configured at APs, and APs and MSs are sparsely distributed. However, if APs and MSs are densely distributed, extra pre-processing methods should be implemented to limit the dimension of node feature.     
\end{remark}

\begin{figure*}
\centering
\includegraphics[width=0.95\linewidth]{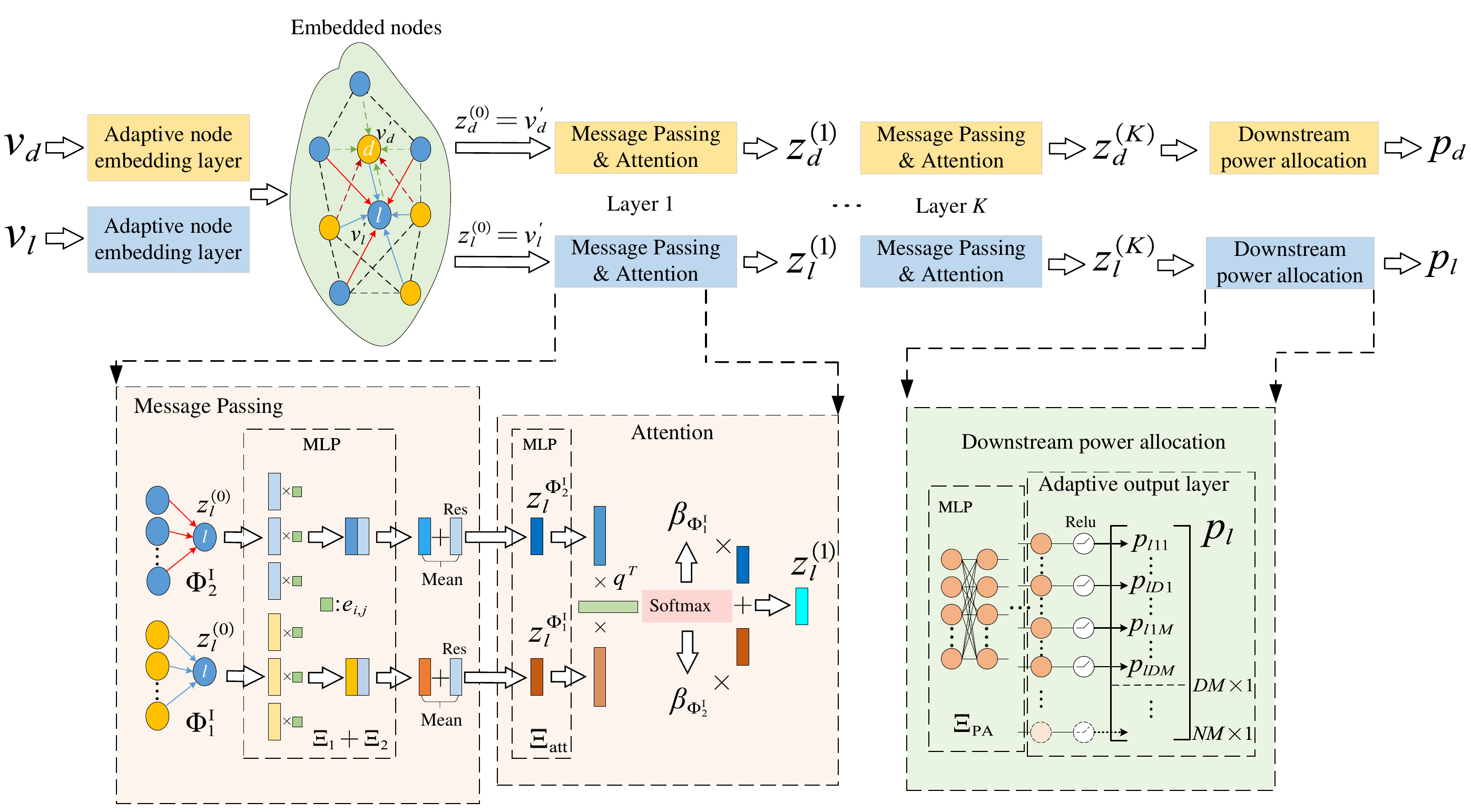}
\caption{The overall architecture of CF-HGNN network.}
\label{figure-MDDCGNN-net}
\end{figure*}

\subsubsection{Meta-Path Based Message Passing}
According to the example shown in Fig. \ref{figure-MDDCGNN-model}, each AP/MS node connects with its neighboring nodes via two meta-paths, and all the meta-paths are mutually independent. Hence, within each meta-path $\Phi$, the process of message passing from neighboring nodes to itself is formulated as
\begin{equation}\label{eq:MDDGNN:MP}
\pmb{z}_i^{\Phi} = \text{Mean}\Big(\Xi_2\Big(\pmb{v}_i^{'}\oplus \frac{1}{\left|\mathcal{N}_i^\Phi\right|}\Xi_1(\sum_{j\in \mathcal{N}_i^\Phi} \pmb{v}_j^{'} \cdot e_{i,j})\Big) + \underbrace{\pmb{v}_i^{'}}_{\text{Res}}\Big),
\end{equation}
where $\Xi_1$ and $\Xi_2$ represent two different MLPs, with each including the fully-connected, activation and normalization layers. The Res $\pmb{v}_i^{'}$ term denotes the residual connection, which can maintain the original node information after the multi-layer message passing. $\oplus$ is the concatenation operation. 

\subsubsection{Meta-Path Based Attention}
In general cases, the information update of node $i$ is the summation of the message collected from all the meta-paths with the same coefficients, expressed as $\pmb{z}_i=\sum_{{\Phi}}\pmb{z}_i^{\Phi}$. However, in MDD-CF networks, since AP/MS nodes receive information via both the interfering path and data transmission path, it is intuitive that these two meth-paths may has quite different impact on the information integration. For instance, if an AP node is closely surrounded by MS nodes but the other AP nodes are far away from it, the meta-path $\Phi_1^{\text{I}}$ should be more important than $\Phi_2^{\text{I}}$. Based on this observation, we propose the meta-based attention in CF-HGNN to enable the GNN to automatically learn the importance of the two meta-paths. An example of calculating the attention vector of an AP node is as follows \cite{velivckovic2017graph}:
\begin{align}\label{eq:MDDGNN:att}
\alpha_{\Phi_1^{\text{I}}} &= \frac{1}{L}\sum_{l\in\mathcal{L}} \pmb{q}^T\Xi_{\text{att}}(\pmb{z}_l^{\Phi_1^{\text{I}}}), \nonumber \\
\beta_{\Phi_1^{\text{I}}} &= \frac{\text{exp}(\alpha_{\Phi_1^{\text{I}}})}{\text{exp}(\alpha_{\Phi_1^{\text{I}}})+\text{exp}(\alpha_{\Phi_2^{\text{I}}})},
\end{align}
where $\Xi_{\text{att}}$ is the MLP layer for attention, $\pmb{q}$ is the learnable attention vector, $\pmb{z}_l^{\Phi_1^{\text{I}}}$ denotes the aggregated information via meta-path $\Phi_1^{\text{I}}$. Then, the final node representation of AP $l$ is $\pmb{z}_l =\beta_{\Phi_1^{\text{I}}}\pmb{z}_l^{\Phi_1^{\text{I}}}+\beta_{\Phi_2^{\text{I}}}\pmb{z}_l^{\Phi_2^{\text{I}}}$, where $\pmb{z}_l\in \mathbb{R}^{F_{\text{AP}}^{'}\times 1}$. Similarly, $\pmb{z}_d\in \mathbb{R}^{F_{\text{MS}}^{'}\times 1}$ at MS $d$ can be obtained. It is noteworthy that the process of message passing plus attention can be iteratively implemented for $K$ times by initializing $\pmb{z}_l^{(0)}=\pmb{v}_l^{\prime}$ and $\pmb{z}_d^{(0)}=\pmb{v}_d^{\prime}$, so as to collect the high-hop neighbors. Owing to this, our proposed model can also be termed as the K-layer CF-HGNN.

\subsubsection{Downstream Power Allocation Learning}
After the $K$-th iteration, the final representation of AP and MS nodes, i.e., $\pmb{z}_l^{(K)}$ and $\pmb{z}_d^{(K)}$, are used for the downstream PA learning, which can be expressed as:
\begin{equation}\label{eq:MDDGNN:DP}
\begin{split}
\pmb{p}_l \in \mathbb{R}^{DM\times 1}  &= \text{Relu}\Big(\Xi_{\text{PA}}(\pmb{z}_l^{(K)})^{(1:DM)}\Big), \\
\pmb{p}_d \in \mathbb{R}^{\bar{M}\times 1} &= \text{Relu}\Big(\Xi_{\text{PA}}(\pmb{z}_d^{(K)})\Big),
\end{split}
\end{equation}
where $\Xi_{\text{PA}}$ denotes the MLP for PA learning, and Relu($\cdot$) is used to constrain the power to be positive. Corresponding to the adaptive node embedding considered in part 1), in order to make it feasible for the network with different number of MSs, the adaptive output layer can support the PA for up to $NM$ elements by activating or deactivating the neurons. By contrast, for UL PA, provided that the number of available UL subcarriers is fixed, the size of the output layer for MS node remains to $\bar{M}$, as each MS only needs to allocate UL power. 

The overall architecture of the CF-HGNN for an AP node is shown in Fig. \ref{figure-MDDCGNN-net}, which is the same for MS nodes, except the different size of output layer as above-mentioned. Note that all the MLPs shown in Fig. \ref{figure-MDDCGNN-net} are AP-specific, which means that the MLPs for MSs are different from that for APs.

Finally, to perform the CF-HGNN training in an unsupervised way, we define the loss function as 
%\begin{figure*}[!t]
%\hrulefill
\begin{align}\label{eq:MDDGNN:losstrain}
\mathcal{L}(\pmb{\theta}) &=  \mathbb{E}\bigg[-\Lambda_{\text{SE}}+\sum_{d=1}^{D}\bigg(\kappa_1\text{Relu}\big(\chi_{\text{DL}}-\sum_{m \in \mathcal{M}}\ln(1+\text{SINR}_{d,m})\big) \nonumber \\
&+\kappa_2\text{Relu}\big(\chi_{\text{UL}}-\sum_{\bar{m} \in \mathcal{\bar{M}}}\ln(1+\text{SINR}_{d,\bar{m}})\big) \nonumber \\
&+ \kappa_3\text{Relu}\big(\sum_{\bar{m} \in \bar{\mathcal{M}}}p_{d\bar{m}}-P_d\big)\bigg) + \sum_{l=1}^L\kappa_4\text{Relu}\big(\sum_{m \in \mathcal{M}}\sum_{d \in \mathcal{D}} p_{ldm}-P_l\big)\bigg],
\end{align}
%\vspace*{-6pt}
%\end{figure*}
where $\pmb{\theta}$ denotes all the parameters of the neural network, $\kappa_i$ denotes the weighted factor, and the expectation is taken with respect to the channel realizations. In \eqref{eq:MDDGNN:losstrain}, each ReLU penalty term has a positive value, only if the DL/UL QoS requirements and transmit power budgets are not satisfied. This will enforce the training process towards satisfying the given requirements. Additionally, in \eqref{eq:MDDGNN:losstrain}, the positive parameters $\kappa_i$ give different priorities to the penalty terms.

\section{Simulation Results and Analysis}\label{sec:MDDGNN:sim}
Let us now demonstrate the achievable SE performance of the distributed MDD-CF systems supported by our proposed CF-HGNN.

\subsection{Simulation Setup}
The large-scale fading model is given by \cite{demir2021foundations}:
\begin{equation}
\beta[\text{dB}] = -30.5-36.7\log_{10}{(d)}+\sigma_{\text{sh}}z,
\end{equation}
where $d$ denotes the distance between any two nodes, $\sigma_{\text{sh}}z$ is the shadowing fading with a standard deviation of $\sigma_{\text{sh}}=4$ dB and $z \sim \mathcal{N}(0,1)$.
Furthermore, we assume that APs are capable of providing 30 dB of IAI suppression in the propagation/analog domain by employing the existing approaches \cite{kolodziej2019band}. Then, assume that the 12-bit ADCs are applied, the MDD system can suppress IAI up to 72 dB (i.e., $\xi_l^{\text{MDD-IAI}}=-72$ dB, $\forall l$, in \eqref{eq:MDD_CF:ylbarm}), of which 42 dB is attributed to the digital-domain cancellation by FFT \footnote{The 12-bit ADC has a maximum dynamic range of 42dB, which means it can accommodate up to 42 dB of power of IAI, and transform it into digital signal without extra quantization noise. Then, in the digital domain, since the interference signal is transmitted over DL subcarriers, which is mutually orthogonal to the desired UL signal, MDD systems supported by the FFT operation can then remove the remaining IAI.}. By contrast, as MSs are of lightweight equipments with single antenna and can hardly share channel knowledge with other MSs, they can not actively suppress IMI. However, similar to APs, MSs are able to cancel 42 dB IMI in the digital domain (i.e., $\xi_d^{\text{MDD-IMI}}=-42$ dB, $\forall d$, in \eqref{eq:MDD_CF:ydm}) with the aid of the FFT operation \cite{day2012full2}. Unless otherwise noted, the simulation parameters of the MDD-CF network are listed in Table \ref{Table:MDDGNN:para}.

As for the settings of the neural network, we adopt a 2-layer CF-HGNN based on Pytorch Geometric \cite{fey2019fast}. The general MLPs $\Xi_1$ and $\Xi_2$ in \eqref{eq:MDDGNN:MP} during the message passing stage contains multiple fully-connected linear layers followed by the LeakyRelu activation layer and batch normalization layer. By contrast, the $\Xi_{\text{PA}}$ in \eqref{eq:MDDGNN:DP} is employed with multiple fully-connected linear layers without batch normalization layer. Moreover, the $\Xi_{\text{ada}}$ in \eqref{eq:MDDGNN:embed} is a single-layer MLP with one fully-connected linear layer followed by one batch normalization layer, while $\Xi_{\text{att}}$ in \eqref{eq:MDDGNN:att} is a single-layer MLP with only one fully-connected linear layer. As formulated in \eqref{eq:MDDGNN:losstrain}, the overall learning is unsupervised without any ground truth. To optimize the CF-HGNN, we adopt the Adam optimizer with a learning rate of 0.001 \cite{kingma2014adam}. Furthermore, we empirically set $\kappa_i$ in \eqref{eq:MDDGNN:losstrain} as $\{0.1,1,0.1,0.1\}$ during training. For the training data, we randomly generate 10000 and 1000 CF network layout samples for training and testing, respectively, under the assumption that APs and MSs are uniformly distributed within a square area of $(S_D \times S_D) \text{m}^2$ . The batch size for training is 64, and the network parameters are only updated during training, which stay constant during testing. The CF-HGNN is run on a GeForce GTX laptop 3080Ti, while the other algorithms are implemented on the 12th Gen Intel(R) Core(TM) i7-12700H 2.70 GHz.

For comparison, we introduce QT-SCA described in Appendix \ref{app:MDDGNN:QTSCA} as a benchmark, which is implemented using the CVX tool during simulations \cite{cvx}. In addition, our proposed CF-HGNN is also compared with the greedy unfair allocation method \cite{jang2003transmit}, where the water-filling algorithm is assumed by each AP/MS to distribute their power over the DL/UL subcarriers, regardless of the QoS constraints.

\begin{table}
\caption{Simulation parameters}
\centering
\begin{tabular}{l|l}
\hline
Default parameters & Value  \\ \hline
%Number of APs, MSs ($L,D$)  & (24, 6)  \\ \hline
%AP density & $150 \ \text{AP}/\text{km}^2$ \\ \hline
Cell area ($S_{\text{D}}\times S_{\text{D}}$) & $(400\times 400) \text{m}^2$  \\ \hline
Number of antennas per AP ($N$) & 8 \\ \hline
Number of DL/UL subcarriers ($M,\bar{M}$) & (4, 2) \\ \hline
AP and MS power ($P_l, P_d, \forall l,d$) & $(40, 30) \ \text{dBm}$ \\ \hline
QoS requirements ($\chi_{\text{DL}}, \chi_{\text{UL}}$) & (0.5, 0.1) nats/s/Hz \\ \hline
Noise power ($\sigma^2$)  & -94 dBm   \\ \hline
Delay taps ($U$) & 4 \\ \hline 
%Large-scale fading ($\beta$) & \cite{demir2021foundations} \\ \hline
Residual SI level at AP ($\xi_l^{\text{SI}}, \forall l$) & -120 dB \\ \hline 
Residual SI level at MS ($\xi_d^{\text{SI}}, \forall d$) & -110 dB \\ \hline 
\end{tabular}
\label{Table:MDDGNN:para}
\end{table}

\subsection{Performance Comparison}
In this section, we make a comprehensive performance comparison between different PA methods in MDD-CF networks, where the numbers of MSs and APs are fixed during training and testing, and are set to $L=24, D=6$. In this case, the embedding matrices in $\Xi_{\text{ada}}$ and $\Xi_{\text{PA}}$ act as identical matrices and are not learnable. 

\begin{figure}
\centering
\includegraphics[width=0.5\linewidth]{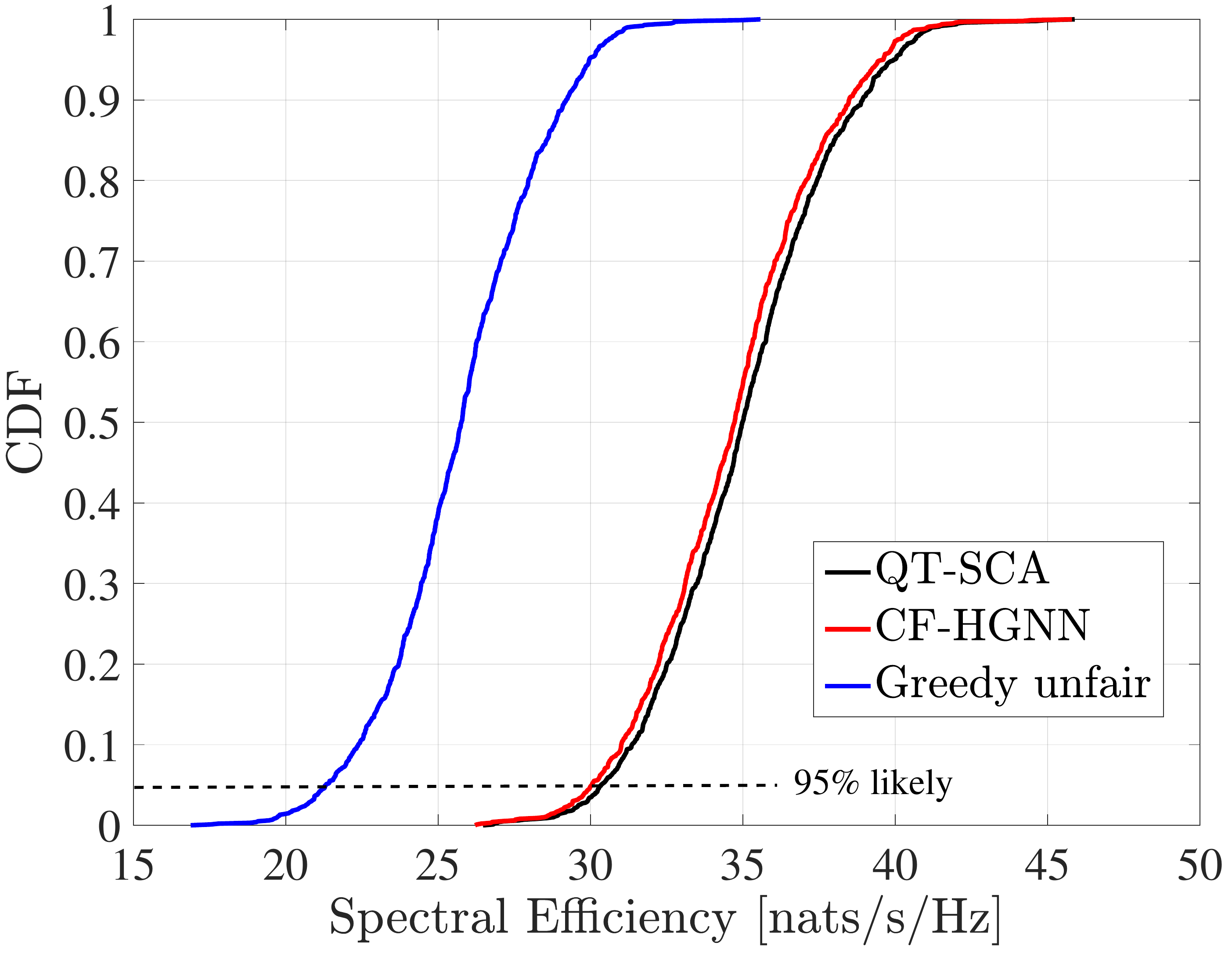}
\caption{Cumulative distribution of SE, when the MDD-CF network has $L=24$ APs and $D=6$ MSs.}
\label{figure-MDDCGNN-sim-layout}
\end{figure}

\begin{figure}
\centering
\includegraphics[width=0.5\linewidth]{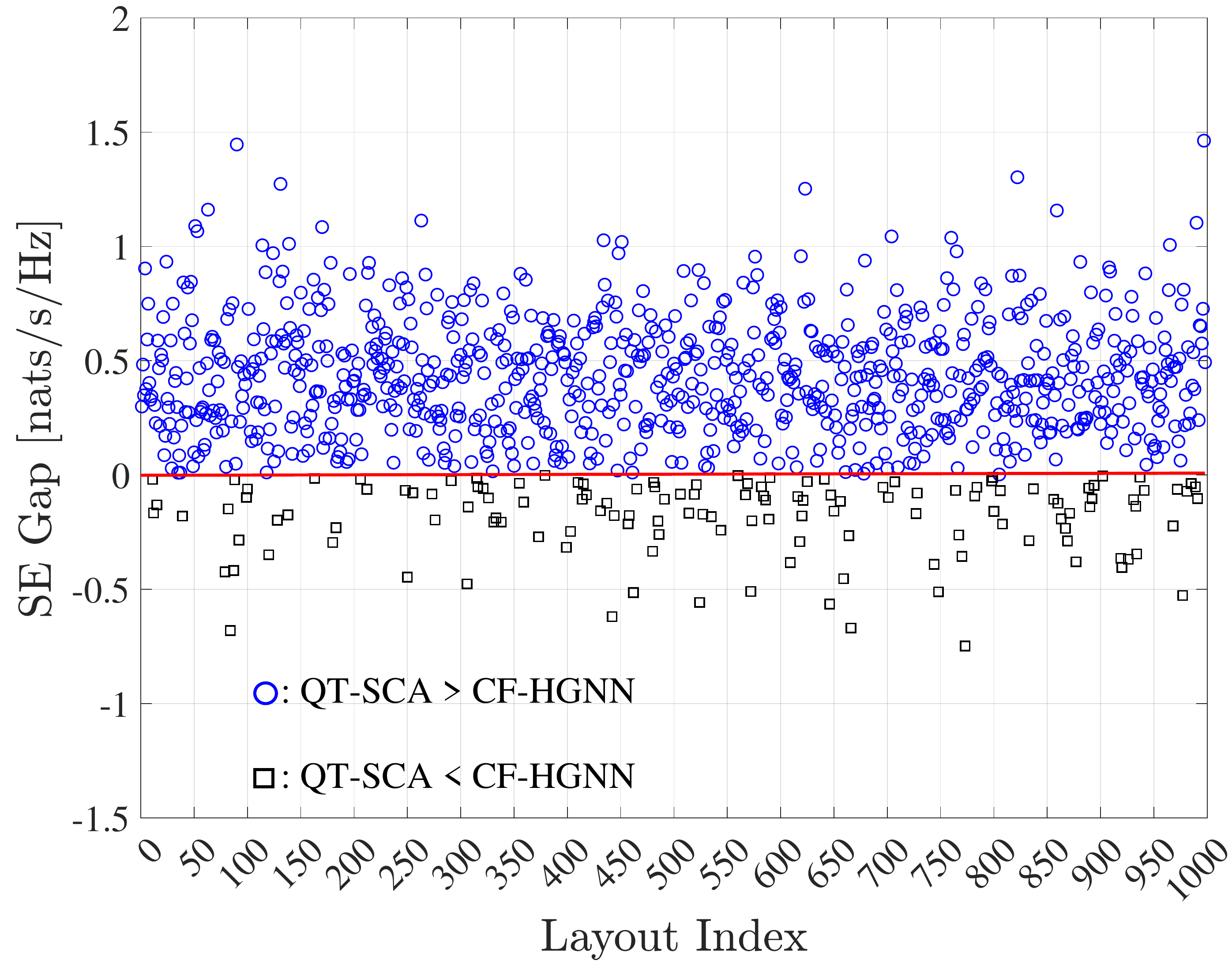}
\caption{SE performance gap between QT-SCA and CF-HGNN, when the MDD-CF network uses $L=24$ APs to support $D=6$ MSs.}
\label{figure-MDDCGNN-sim-gap}
\end{figure}

Firstly, we compare the QT-SCA, CF-HGNN and the greedy unfair methods in terms of the SE distribution obtained from 1000 testing CF network layouts.  As shown in Fig. \ref{figure-MDDCGNN-sim-layout}, the CF-HGNN achieves nearly the same performance as the QT-SCA in terms of the 95\%-likely SE. In more detail, the SE performance gaps between the QT-SCA and CF-HGNN with regard to these 1000 network layouts are depicted in Fig. \ref{figure-MDDCGNN-sim-gap}. It can be observed that the absolute SE gaps between these two methods are lower than 1.5 nats/s/Hz. Furthermore, there are several layouts, where the CF-HGNN outperforms the QT-SCA, which reflects that the CF-HGNN is capable of learning the near-optimal solutions for PA in MDD-CF networks. As shown in Fig. \ref{figure-MDDCGNN-sim-layout}, both the QT-SCA and CF-HGNN significantly outperform the greedy unfair method. The rationale behind is that although the greedy unfair method can maximize the SE of the classic HD-based multiuser OFDM systems \cite{jang2003transmit}, it lacks the capability to manage the complicated interference in the CF FD-like systems, hence leading to poor performance at both APs and MSs.

\begin{figure}
\centering
\includegraphics[width=0.5\linewidth]{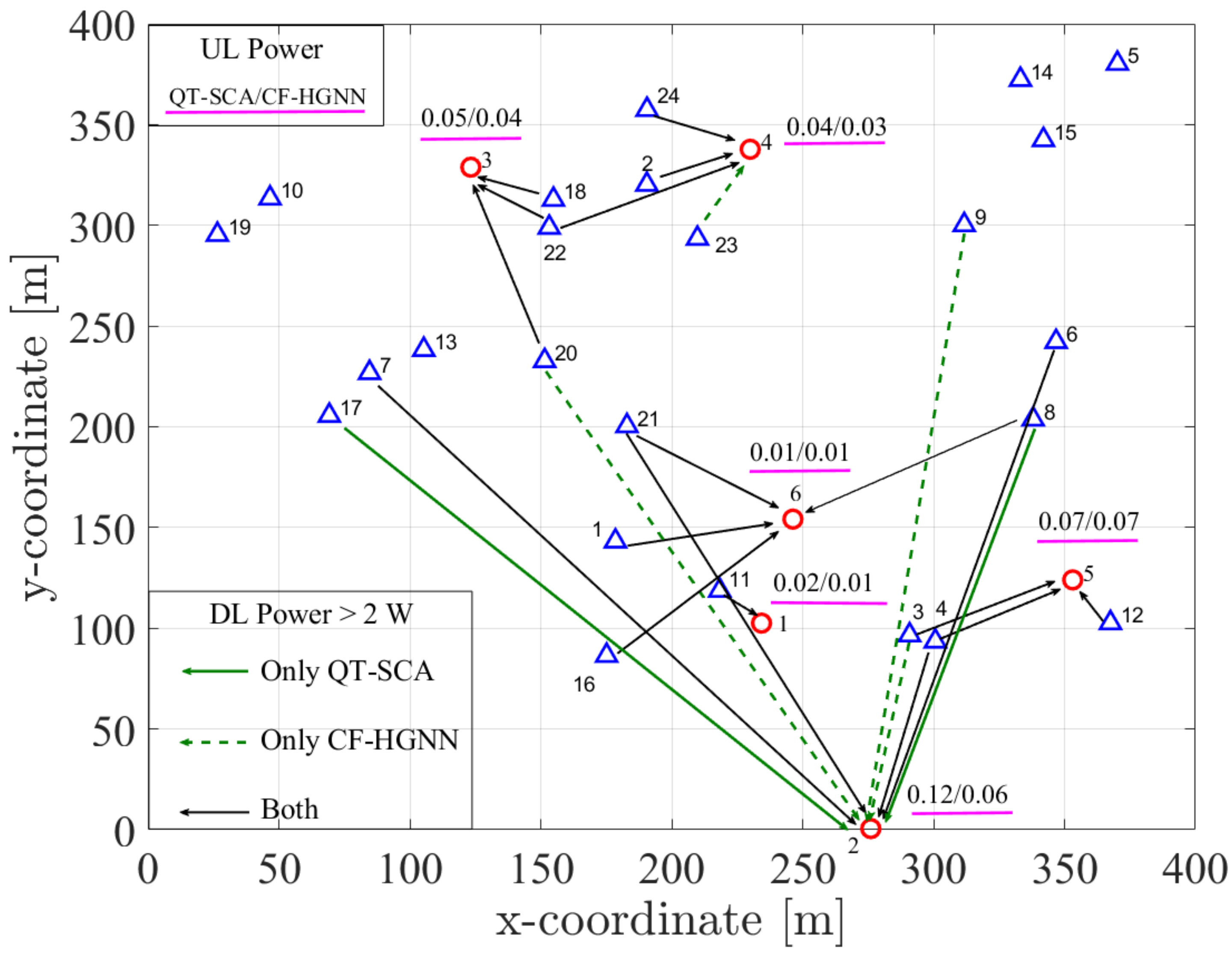}
\caption{An MDD-CF network topology with $L=24$ APs and $D=6$ MSs.}
\label{figure-MDDCGNN-sim-Connect}
\end{figure}

Secondly, we randomly select one of the CF network layouts used in the testing samples, as shown in Fig. \ref{figure-MDDCGNN-sim-Connect}, to deeply investigate the PA attained by the QT-SCA and CF-HGNN. To make the drawing clear in Fig. \ref{figure-MDDCGNN-sim-Connect}, the DL connections with transmit power less than $2 W$ are omitted, and different colored lines are used to denote the DL connections obtained by either methods or by both. It is not surprise that, as Fig. \ref{figure-MDDCGNN-sim-Connect} shows, except MS 2, both the CF-HGNN and QT-SCA yield the same subset of the major serving APs for each of MSs, and also obtain the similar results of UL PA. As for MS 2, since it locates relatively far away from APs, more APs are required to transmit signal to it in order to meet the demand of DL's QoS. Although different DL connections are obtained by the two methods, there is in fact no difference between them. For example, apart from the common connections, the CF-HGNN picks AP 3 and 9 to serve MS 2, while the QT-SCA chooses AP 8 located between AP 3 and 9 to serve MS 2. Also, AP 20 and 17 with the similar distances from MS 2 are selected by the CF-HGNN and QT-SCA, respectively. 

It is worth mentioning that the DL power (and UL power) used for Fig. \ref{figure-MDDCGNN-sim-Connect} is the sum of power allocated to DL subcarriers (and UL subcarriers). In Table \ref{Table:MDDGNN:power}, an example showing the detailed PA to subcarriers at AP 1, MS 2 and MS 6 are presented. It can be observed that the CF-HGNN and greedy unfair methods yield the more consistent allocation among the subcarreirs, which means that the power distribution is largely relied on large-scale fading, while the QT-SCA has better capability of exploiting the small-scale fading. Moreover, from the results of the CF-HGNN and QT-SCA, we can see that MS 6 has a lower UL transmit power than MS 2. The reason behind is that MS 6 is in the close proximity of MS 1, the increase of its UL transmit power may cause not only more SI on the DL reception, but also larger IMI on MS 1, leading to the degradation of SE. By contrast, the greedy unfair scheme allocates power depending only on the quality of communication channels regardless of the effect of interference.

\begin{table*}[!t]
\caption{PA among Subcarriers  at AP 1, MS 2 and MS 6.}
\centering
\begin{tabular}{|c|c|c|c|c|c|c|c|c|c|}
\hline
\multicolumn{1}{|c|}{\multirow{2}{*}{Connections}} &\multirow{2}*{Method} & \multicolumn{4}{c|}{\multirow{2}{*}{\makecell[c]{DL subcarriers' power \\ ($W$)}}} & \multicolumn{1}{c|}{\multirow{2}{*}{\makecell[c]{DL power \\ ($W$)}}} & \multicolumn{2}{c|}{\multirow{2}{*}{\makecell[c]{UL subcarriers' power \\ ($W$)}}} & \multicolumn{1}{c|}{\multirow{2}{*}{\makecell[c]{UL power \\ ($W$)}}} \\ 
\multicolumn{1}{|c|}{}&\multicolumn{1}{c|}{} & \multicolumn{4}{c|}{} & \multicolumn{1}{c|}{} & \multicolumn{2}{c|}{} & \multicolumn{1}{c|}{} \\ \hline
\multirow{2}{*}{AP 1$\rightarrow$MS 6,}&QT-SCA & 1.89 & 0.81 & 1.90 & 1.89 & \textbf{6.49} & 0.006 & 0.008 & \textbf{0.014}\\ 
\cline{2-10}
\multirow{1}{*}{}&CF-HGNN & 1.77 & 1.74 & 1.78 & 1.79 & \textbf{7.08} & 0.003 & 0.010 & \textbf{0.013} \\ 
\cline{2-10}
MS 6$\rightarrow$APs&Greedy unfair & 1.26 & 1.19 & 1.25 & 1.25 & \textbf{4.95} & 0.499 & 0.501 & \textbf{1.000} \\ \hline
\multirow{2}{*}{AP 1$\rightarrow$MS 2,}&QT-SCA & 0.75 & 0.33 & 0.49 & 0.31 & \textbf{1.88} & 0.026 & 0.038 & \textbf{0.064}\\ 
\cline{2-10}
\multirow{1}{*}{}&CF-HGNN & 0.25 & 0.23 & 0.22 & 0.27 & \textbf{0.97} & 0.000 & 0.117 & \textbf{0.117} \\ \cline{2-10}
MS 2$\rightarrow$APs&Greedy unfair & 0.00 & 0.00 & 0.00 & 0.00 & \textbf{0.00} & 0.000 & 1.000 & \textbf{1.000} \\ \hline
\end{tabular}
\label{Table:MDDGNN:power}
\end{table*}

\begin{figure}
\centering
\includegraphics[width=0.5\linewidth]{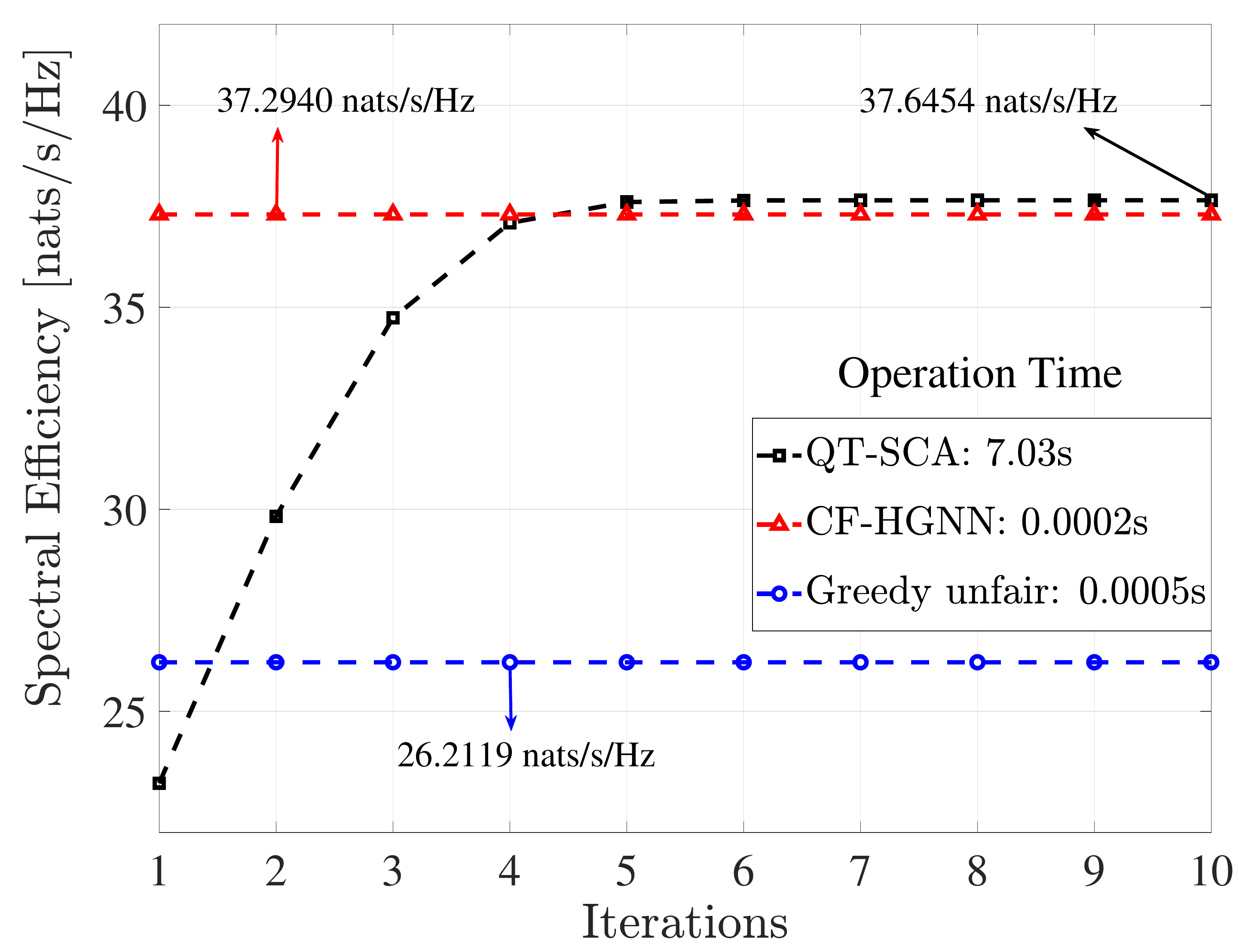}
\caption{SE convergence behavior and operation time of different methods, where $L=24, D=6$.}
\label{figure-MDDCGNN-sim-converge}
\end{figure}

The performance of different methods in terms of SE, convergence behavior and operation time is plotted in Fig. \ref{figure-MDDCGNN-sim-converge}. As expected, the QT-SCA and CF-HGNN achieve the comparable SE, with a performance gap of only 0.35 nats/s/Hz. However, the QT-SCA is an iterative algorithm with very high computational complexity, which converges within 6 iterations and takes 7.03s operation time in total. On the contrary, with the aid of the parallel computation of GPU, the CF-HGNN is capable of reaching 99\% of the SE achieved by the QT-SCA but using only $10^{-4}$ times of its operation time to. As to the greedy unfair method, although it has a similar operation time as the CF-HGNN, due to the fact that only the water-filling algorithm is applied at both APs and MSs, its SE performance is significantly worse than that of the other two methods.

Next, we evaluate the generalization performance of the proposed CF-HGNN in MDD-CF networks with different cell sizes. As shown in Table \ref{Table:MDDGNN:vscellsize}, the results under the Generalized CF-HGNN (GCF-HGNN) are obtained by the CF-HGNN model trained on the specific network with $S_{\text{D}}=400$, while the results under the Dedicated CF-HGNN (DCF-HGNN) are obtained by the CF-HGNN models trained correspondingly using different cell sizes. It can be observed from the table that as the cell size reduces, the performance of the GCF-HGNN degrades slightly, but is still in the acceptable range. Note that, the DCF-HGNN finally outperforms the QT-SCA when $S_{\text{D}}=150$, which means that the DCF-HGNN is capable of obtaining a better PA strategy in dense networks.

\begin{table*}[!t]
\caption{Generalization to Different Cell Sizes. The SE of the CF-HGNN are normalized by that of the QT-SCA.}
\centering
\begin{tabular}{|c|c|c|c|}
\hline
\multirow{2}*{$S_{\text{D}}$} & \multirow{2}*{\makecell[c]{QT-SCA \\ (nats/s/Hz)}}& \multirow{2}*{DCF-HGNN} & \multirow{2}*{GCF-HGNN}  \\ 
\multicolumn{1}{|c|}{} & \multicolumn{1}{c|}{} & \multicolumn{1}{c|}{} & \multicolumn{1}{c|}{} \\ \hline
%400 & 35.02 & 97.66\% &97.66\% \\ \hline
350 & 37.85 &96.64\%  &95.19\% \\ \hline
300 & 40.64 &98.08\%  &93.13\% \\ \hline
250 & 43.76 &98.42\%  &91.84\%  \\ \hline
200 & 46.84 &99.91\%  &91.78\%  \\ \hline
150 & 50.10 &101.42\%  &92.79\% \\ \hline
%100 & 56.04 &101.42\%  &91.08\% \\ \hline
\end{tabular}
\label{Table:MDDGNN:vscellsize}
\end{table*}

\begin{figure}
\centering
\includegraphics[width=0.5\linewidth]{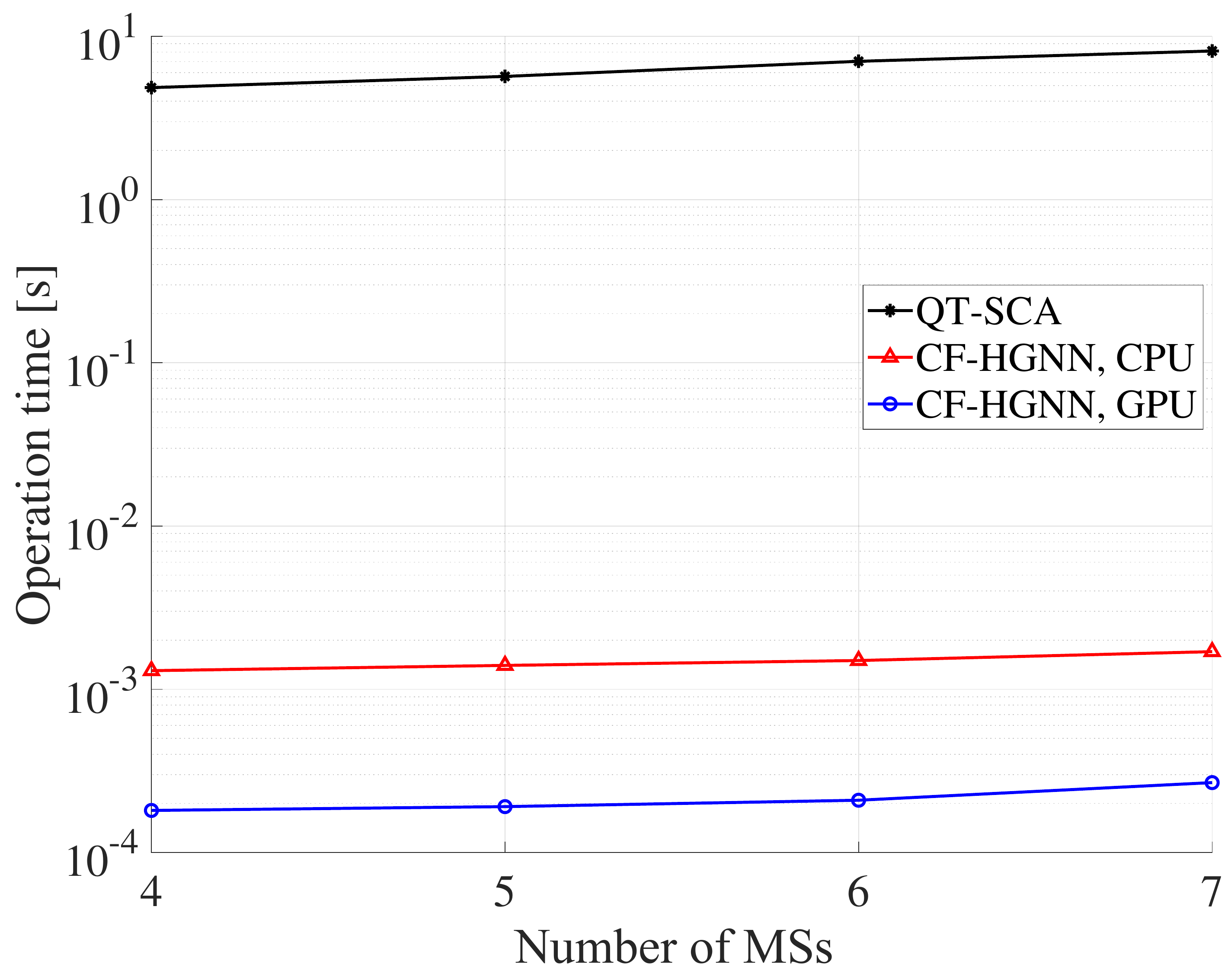}
\caption{Comparison of operation time between the QT-SCA and CF-HGNN methods.}
\label{figure-MDDCGNN-sim-time}
\end{figure}

The comparison between QT-SCA and CF-HGNN in terms of computational complexity and operation time are then analyzed. According to \cite{peaucelle2002user}, the approximated computational complexity of the QT-SCA method is $\mathcal{O}\big((LDM+2DM+3D\bar{M})^2(L+D+3DM_{\text{sum}})^{2.5}+(L+D+3DM_{\text{sum}})^{3.5}\big)$ per iteration. As to the CF-HGNN, the computational complexity is mainly due to the matrix computation, as shown in \eqref{eq:MDDGNN:MP}-\eqref{eq:MDDGNN:DP}. Table \ref{Table:MDDGNN:com} summarizes the results for the different values of $L,D,M$ and $\bar{M}$, showing that when doubling the numbers of APs, MSs and subcarriers, the computational complexity of the QT-SCA increases much more quickly than that of the CF-HGNN, which is about three orders for the QT-SCA versus one order for the CF-HGNN. The operation time cost by the QT-SCA and CF-HGNN is plotted in Fig. \ref{figure-MDDCGNN-sim-time}. Explicitly, the CF-HGNN spends much less time than the QT-SCA to accomplish the PA. Moreover, thanks to the parallel computation of GPU, the CF-HGNN trained on GPU has the lowest operation time. 

\begin{table}[!t]
\caption{Computational complexity comparison}
\centering
\begin{tabular}{|c|c|c|c|c|}
\hline
\multirow{2}*{$L$} &\multirow{2}*{$D$} & \multirow{2}*{$M/\bar{M}$} & \multicolumn{2}{c|}{Methods}  \\ \cline{4-5}
\multicolumn{1}{|c|}{} & \multicolumn{1}{c|}{} & \multicolumn{1}{c|}{} & \multicolumn{1}{c|}{QT-SCA} & \multicolumn{1}{c|}{CF-HGNN} \\ \hline
6 & 2 & 4/2 & $4.48\times 10^{8}$ & $6.83\times 10^{6}$ \\ \hline
12 & 4 & 8/4 & $4.78\times 10^{11}$ & $2.75\times 10^{7}$ \\ \hline
24 & 8 & 16/8 & $6.79\times 10^{14}$ & $1.48\times 10^{8}$ \\ \hline
%20& \multirow{4}{*}{$6$}&\multirow{4}{*}{$4/2$} & $3.97\times 10^{11}$ & $2.39\times 10^{7}$ \\ 
%\cline{1-1}\cline{4-5}
%21&\multirow{1}{*}{}&\multirow{1}{*}{} & $4.39\times 10^{11}$ & $2.48\times 10^{7}$ \\  
%\cline{1-1}\cline{4-5}
%22&\multirow{1}{*}{}&\multirow{1}{*}{} & $4.84\times 10^{11}$ & $2.58\times 10^{7}$ \\ 
%\cline{1-1}\cline{4-5}
%23&\multirow{1}{*}{}&\multirow{1}{*}{} & $5.33\times 10^{11}$ & $2.67\times 10^{7}$ \\  \hline
%\multirow{4}{*}{$24$}&4 & \multirow{4}{*}{$4/2$} &$1.16\times 10^{11}$ &$2.49\times 10^{7}$ \\ \cline{2-2} \cline{4-5}
%\multirow{1}{*}{}&5 &\multirow{1}{*}{} &$2.82\times 10^{11}$ &$2.62\times 10^{7}$ \\ \cline{2-2} \cline{4-5}
%\multirow{1}{*}{}&6 &\multirow{1}{*}{} & $5.85\times 10^{11}$ &$2.77\times 10^{7}$  \\ \cline{2-2} \cline{4-5}
%\multirow{1}{*}{}&7 &\multirow{1}{*}{} & $1.09\times 10^{12}$ &$2.91\times 10^{7}$  \\ \hline
%\multirow{4}{*}{24} & \multirow{4}{*}{6} & 8/4 & 9.26 & 98.81\\ \cline{3-3}\cline{4-5}
%\multirow{1}{*}{} & \multirow{1}{*}{} & 16/8 & 15.53 & 99.03 \\ \hline
\end{tabular}
\label{Table:MDDGNN:com}
\end{table}

\subsection{Scalability: Case 1} 
The simulation results in Section IV-B demonstrated that the CF-HGNN is able to achieve the similar performance as the QT-SCA. However, the CF-HGNN considered so far can only be applied to the specific MDD-CF networks with the fixed numbers of nodes and subcarriers, as the fully-connected liner layers in $\Xi_{\text{ada}}$ and $\Xi_{\text{PA}}$ act as identical matrices. With this regard, to enable the CF-HGNN to be adaptive to the varying MDD-CF networks, we now invoke the adaptive node embedding layer and adaptive output layer. In particular, we consider two cases regarding to the scalability of CF-HGNN. In Case 1, we still assume that $L<L^{\prime}=24$ and $N=8$, and all MSs are served by all APs using the ZF beamforming under the constraint of $N\geq D$. Specifically, 10000 MDD-CF network samples with the specific numbers of nodes and subcarriers are leveraged to train the DCF-HGNN. For fair comparison, to train the Adaptive CF-HGNN (ACF-HGNN), 10000 MDD-CF network samples consisting of various numbers of nodes and subcarriers are used. 

Table \ref{Table:MDDGNN:vsslim} shows that by employing the adaptive layers, the ACF-HGNN can attain the relatively stable performance, when dealing with the various networks. Hence, it is feasible for implementing dynamic PA in MDD-CF networks. On the contrary, the QT-SCA and DCF-HGNN have to re-solve the complicated optimization problem or re-train the CF-HGNN, once the CF network has some changes in terms of the numbers of APs, MSs and subcarriers. Note that, in the table, the SE decreases with the increase of the number of subcarriers. This is because the total power allocated to subcarriers is constrained, while the SE is normalized by $M_{\text{sum}}$, as shown in \eqref{eq:MDDGNN:SE_formulation}. Apparently, the total SE obtained by multiplying $M_{\text{sum}}$ with the value in the table always increases with the increase of the number of subcarriers, which is attributed to the subcarrier diversity.     
%As 10000 samples are used for training in the fixed network, for fair comparison, we will also use 10000 samples in total consisting of different size of network input to train Adaptive CF-HGNN model.

\begin{table*}[!t]
\caption{Generalization to the different numbers of APs, MSs and subcarriers. The SE of the CF-HGNN and greedy unfair algorithm are normalized by that of the QT-SCA.}
\centering
\begin{tabular}{|c|c|c|c|c|c|c|}
\hline
\multirow{2}*{$L$} &\multirow{2}*{$D$}&\multirow{2}*{$M/\bar{M}$} & \multirow{2}*{\makecell[c]{QT-SCA \\ (nats/s/Hz)}} & \multirow{2}*{DCF-HGNN} & \multirow{2}*{ACF-HGNN}& \multirow{2}*{Greedy unfair} \\ 
\multicolumn{1}{|c|}{}&\multicolumn{1}{c|}{}&\multicolumn{1}{c|}{} & \multicolumn{1}{c|}{} & \multicolumn{1}{c|}{} & \multicolumn{1}{c|}{} & \multicolumn{1}{c|}{} \\ \hline
\multirow{4}{*}{$24$}& 4 & \multirow{4}{*}{$4/2$} & 26.42 & 98.07\% & 94.14\% & 74.45\%\\
\cline{2-2}\cline{4-7}
\multirow{1}{*}{}&5&\multirow{1}{*}{} & 31.34 & 98.02\% & 94.51\%& 74.06\%\\ 
\cline{2-2}\cline{4-7}
\multirow{1}{*}{}&6&\multirow{1}{*}{} & 34.23 & 97.78\%  & 94.77\% & 73.06\%\\ 
\cline{2-2}\cline{4-7}
\multirow{1}{*}{}&7&\multirow{1}{*}{}  & 36.24 & 98.18\%& 95.45\% & 73.68\%\\ \hline
20& \multirow{4}{*}{$6$}&\multirow{4}{*}{$4/2$} & 33.04 & 98.09\%  & 97.22\%& 72.79\%\\ 
\cline{1-1}\cline{4-7}
16&\multirow{1}{*}{}&\multirow{1}{*}{} & 29.99 & 97.70\% & 95.07\% & 71.16\%\\ 
\cline{1-1}\cline{4-7}
12&\multirow{1}{*}{}&\multirow{1}{*}{} & 25.69 & 98.52\%  & 94.08\% & 69.95\%\\ 
\cline{1-1}\cline{4-7}
8&\multirow{1}{*}{}&\multirow{1}{*}{}  & 21.13 & 97.40\% & 94.46\%& 68.67\% \\ \hline
\multirow{2}{*}{16} & \multirow{2}{*}{6} & 32/16 &9.26 & 98.81\%  & 91.58\% &72.35\% \\ \cline{3-7}
\multirow{1}{*}{} & \multirow{1}{*}{} & 16/8 & 15.53 & 99.03\% & 91.50\%& 68.45\%  \\ \hline
\end{tabular}
\label{Table:MDDGNN:vsslim}
\end{table*}

\subsection{Scalability: Case 2} \label{sec:MDD_GNN:case2}
Although the proposed ACF-HGNN can handle the PA in various MDD-CF networks, it can hardly cut the mustard when the numbers of APs and MSs become too large, i.e., when $L\gg L^{\prime}$ and $D\gg N$, as explained in Remark \ref{remark:MDD_GNN:re2}. Therefore, in Case 2, we resort to the user-centric clustering strategy to transform the dense graph into the sparse graph, thereby maintaining the scalability of the CF-HGNN for operation in the large-scale MDD-CF networks. For the sake of explanation, here we consider an extreme scenario, where each AP is only equipped with one antenna and serves only one MS using the ZF beamforming. Correspondingly, the clustering can be achieved in two steps\footnote{Note that the considered two-stage clustering approach is suboptimal, and becomes less efficient in the case of $N>1$, where each AP can simultaneously serve more than one MS using the increased spatial degrees of freedom. To this end, our future work motivates to propose a GNN-assisted deep Q-learning network so as to  obtain the optimal user grouping and PA at the same time.}: 1) Initialize the set of APs serving MS $d$ as $\mathcal{L}_d=\emptyset, \forall d=1,...,D$, and the set of MSs assigned to AP $l$ as $\mathcal{D}_l=\emptyset, \forall l=1,...,L$. To guarantee a non-zero SE, each MS $d$ is firstly assigned to a master AP by following the optimization of
\begin{equation}\label{eq:MDDGNN:ucl}
l=\argmax_{l\in\{1,...,L\},m\in\{1,...,M\}} \omega_{ldm}, \ \   \mathcal{L}_d=\{l\}, \  \mathcal{D}_l=\{d\} \ .
\end{equation}
After this optimization, we have $\left|\mathcal{L}_d\right|=1, \forall d=1,...,D$. 2) If there are still idle APs, which satisfy $\{l^{\prime}|\mathcal{D}_{l^{\prime}}=\emptyset\}$, allocate them to MSs according to the optimization of
\begin{equation}\label{eq:MDDGNN:ucd}
d=\argmax_{d\in\{1,...,D\},m\in\{1,...,M\}} \omega_{l^{\prime}dm}, \ \   \mathcal{D}_{l^{\prime}}=\{d\}, \ \mathcal{L}_d=\mathcal{L}_d\cup\{l^{\prime}\}  \ .
\end{equation}

To train the Large-Scale ACF-HGNN (LSACF-HGNN), we have 10000 training samples collected from the networks with $L=12$ single-antenna APs and $D=6$ MSs uniformly distributed within an area of ($200\sqrt{2}\times200\sqrt{2})\text{m}^2$. We then increase the number of APs and MSs in testing samples while the densities of AP and MSs are fixed. Note that all the involved training and testing samples have been pre-processed by the proposed clustering algorithm. As shown in Table \ref{Table:MDDGNN:vslargescale}, although the performance of the LSACF-HGNN is much better than that of the greedy unfair method, the gap between the QT-SCA and the LSACF-HGNN becomes larger as the network size increases. The rationale behind is possibly because the larger CF network has much more complicated CLI problem than the smaller one. Hence, the LSACF-HGNN may require extra training or an improved model structure to manage the CLI. The detailed analysis is out of the scope of this paper and is left for the future research. However, we should mention that, as shown in Table \ref{Table:MDDGNN:com}, the QT-SCA relied optimization is extremely time-consuming, when it is applied to the large-scale CF networks, while the LSACF-HGNN is capable of accomplishing PA with high-efficiency. 

%Note that, as each AP individually transmit signal only with one antenna, and hence the multiuser interference cannot be fully mitigated. Consequently, there is a feasible solution obtained by QT-SCA, the $\chi_{\text{DL}}, \chi_{\text{UL}}$ are set to 

\begin{table*}[!t]
\caption{Generalization to the Large-scale MDD-CF networks. The SE of the CF-HGNN and greedy unfair algorithm are normalized by that of the QT-SCA.}
\centering
\begin{tabular}{|c|c|c|c|c|c|}
\hline
\multirow{2}*{$S_{\text{D}}$} &\multirow{2}*{$L$} &\multirow{2}*{$D$}& \multirow{2}*{\makecell[c]{QT-SCA \\ (nats/s/Hz)}}& \multirow{2}*{LSACF-HGNN} & \multirow{2}*{Greedy unfair}  \\ 
\multicolumn{1}{|c|}{} & \multicolumn{1}{c|}{}& \multicolumn{1}{c|}{}& \multicolumn{1}{c|}{} & \multicolumn{1}{c|}{} & \multicolumn{1}{c|}{} \\ \hline
%400 & 35.02 & 97.66\% &97.66\% \\ \hline
400 &24 & 12 &22.93 &93.68\%  &65.68\% \\ \hline
450 &30 & 15 & 28.28 &92.33\%  &65.38\% \\ \hline
500 &38 & 19 & 35.18 &91.42\%  &65.66\%  \\ \hline
400$\sqrt{2}$ &48 & 24 & 43.93 &88.05\%  &65.65\%  \\ \hline
400$\sqrt{3}$ &72 & 36 & 63.92 &82.09\%  &65.35\% \\ \hline
800 &96 & 48 & 82.10 &80.58\%  &67.23\% \\ \hline
\end{tabular}
\label{Table:MDDGNN:vslargescale}
\end{table*}

\section{Conclusions}
In this paper, we proposed a distributed MDD-CF system, and investigated the PA at both AP and MS sides for SE maximization under the constraints of QoS requirements. In order to solve the non-convex and NP-hard PA problem, we first proposed the QT-SCA algorithm, which achieves the optimization in the classic way. Then, the CF-HGNN was proposed to solve the optimization from a learning perspective. The CF-HGNN was trained in an unsupervised fashion with unlabeled data, which significantly reduces the system overhead. Our studies and numerical results show that the CF-HGNN is capable of achieving comparable SE performance to the QT-SCA but demanding much less operation time and computational complexity. The CF-HGNN significantly outperforms the greedy unfair method in terms of SE performance. Furthermore, with the aid of adaptive node embedding layer and adaptive output layer, the CF-HGNN can implement PA in the MDD-CF networks with various numbers of APs, MSs and subcarriers. Additionally, with the aid of user clustering, the CF-HGNN trained based on a relatively small-scale MDD-CF network can be generalized for operation in the large-scale MDD-CF networks. 
%In the future research, we aim to integrate the deep Q-learning into the CF-HGNN so as to obtain the optimal user clustering and PA results.  
% so as to mitigate the restriction on the number of MSs, which is assumed to be less than the number of antennas configured at each AP in this paper.

\appendices
\section{QT-SCA Optimization}\label{app:MDDGNN:QTSCA}
As the main function of \eqref{eq:MDDGNN:SE_formulation} belongs to the general multiple-ratio concave-convex fractional programming problem, it can be firstly reformulated by the QT method \cite{shen2018fractional} as
\begin{align}\label{eq:MDDGNN:QT}
&\max_{\pmb{p},\pmb{z}} \frac{1}{M_{\text{sum}}}\sum_{d=1}^D \bigg(\sum_{m=1}^M \ln\big(1+2z_{dm}\sqrt{A_{d,m}(\pmb{p})}-z_{dm}^2B_{d,m}(\pmb{p})\big) \nonumber \\
&+\sum_{\bar{m}=1}^{\bar{M}} \ln(1+2z_{d\bar{m}}\sqrt{A_{d,\bar{m}}(\pmb{p})}-z_{d\bar{m}}^2 B_{d,\bar{m}}(\pmb{p})\bigg),
\end{align}
where $\text{SINR}_{d,m}=\frac{A_{d,m}(\pmb{p})}{B_{d,m}(\pmb{p})}$, $\text{SINR}_{d,\bar{m}}=\frac{A_{d,\bar{m}}(\pmb{p})}{B_{d,\bar{m}}(\pmb{p})}$, $\pmb{p}=(\left\{p_{ldm}\right\},\left\{p_{d\bar{m}}\right\})$. In \eqref{eq:MDDGNN:QT}, $\ln(1+x)$ is a non-decreasing and concave function. For a given $\pmb{p}$, both $\text{SINR}_{d,m}$ and $\text{SINR}_{d,\bar{m}}$ are also concave-convex. Hence, when $\pmb{p}$ is fixed, the optimal $\pmb{z}=\left(\left\{z_{dm}\right\},\left\{z_{d\bar{m}}\right\}\right)$ maximizing \eqref{eq:MDDGNN:QT} can be obtained as $\pmb{z}^{\ast}=\frac{\sqrt{A(\pmb{p})}}{B(\pmb{p})}$. Then, given a fixed value of $\pmb{z}$, the problem \eqref{eq:MDDGNN:QT} is a concave maximization problem over $\pmb{p}$.

However, the constraints of (\ref{eq:MDDGNN:SE_formulation}d) and (\ref{eq:MDDGNN:SE_formulation}e) are still nonconvex, which need to be approximated by the convex ones using the SCA method. As shown in (\ref{eq:MDDGNN:SE_formulation}d) and (\ref{eq:MDDGNN:SE_formulation}e), these original constraints are extremely complicated, which contain the sum of $M$ and $\bar{M}$ nonconvex components, respectively. Hence, we firstly simplify them to the following ones: 
\begin{align}
&~~\text{SINR}_{d,m}\geq e^{\frac{\chi_{\text{DL}}}{M}}-1, \ \forall d \in \mathcal{D}, m \in \mathcal{M}, \label{eq:MDDGNN:SINRsimple1}\\
&~~\text{SINR}_{d,\bar{m}}\geq e^{\frac{\chi_{\text{UL}}}{\bar{M}}}-1, \ \forall d \in \mathcal{D}, \bar{m} \in \mathcal{\bar{M}}. \label{eq:MDDGNN:SINRsimple2}
\end{align}

Then, based on \eqref{eq:MDDGNN:SINRSim}, the simplified constraint \eqref{eq:MDDGNN:SINRsimple1} can be equivalently written as
\begin{subnumcases}{\label{eq:MDDCF:SINRdmEqu}}
&$\text{SINR}_{d,m} \triangleq \varpi^2_{d,m} / \psi_{d,m} \geq e^{\frac{\chi_{\text{DL}}}{M}}-1$, \label{eq:MDDGNN:SINRdmEqu:a} \\
&$0<\varpi_{d,m} \leq \sum_{l \in \mathcal{L}}\sqrt{p_{ldm}}\omega_{ldm}$, \label{eq:MDDGNN:SINRdmEqu:b} \\
&$\psi_{d,m} \geq \xi_d^{\text{SI}}\Theta_{\text{DL}}+\sigma^2$, \label{eq:MDDGNN:SINRdmEqu:c}
\end{subnumcases}
where $\varpi_{d,m}$ and $\psi_{d,m}$ are new variables, while \eqref{eq:MDDGNN:SINRdmEqu:b} and \eqref{eq:MDDGNN:SINRdmEqu:c} are linear constraints. For \eqref{eq:MDDGNN:SINRdmEqu:a}, since the function $f_{\text{ca}}(\varpi_{d,m},\psi_{d,m})\triangleq \varpi^2_{d,m} / \psi_{d,m}$ with $(\varpi_{d,m},\psi_{d,m})\in \mathbb{R}_{++}^2$ is convex, it can be approximated with the aid of the SCA properties as \cite{marks1978general}
\begin{align}
&f_{\text{ca}}(\varpi_{d,m},\psi_{d,m})\nonumber \\ 
&\geq \frac{2\varpi_{d,m}^{(t)}}{\psi_{d,m}^{(t)}}\varpi_{d,m}-\frac{(\varpi_{d,m}^{(t)})^2}{(\psi_{d,m}^{(t)})^2}\psi_{d,m} :=f_{\text{ca}}^{(t)}(\varpi_{d,m},\psi_{d,m}),
\end{align}
where $(\varpi_{d,m}^{(t)},\psi_{d,m}^{(t)})$ is the feasible point obtained at the $t$-th iteration. Consequently, \eqref{eq:MDDGNN:SINRdmEqu:a} can be substituted by the new constraint given by
\begin{equation}\label{eq:MDDGNN:SINRCA}
f_{\text{ca}}^{(t)}(\varpi_{d,m},\psi_{d,m})\geq e^{\frac{\chi_{\text{DL}}}{M}}-1.
\end{equation}

Following the same spirit, \eqref{eq:MDDGNN:SINRsimple2} can be equivalently expressed by the following convex constraints:
\begin{subnumcases}{\label{eq:MDDGNN:SINRdbarmEqu}}
&$f_{\text{ca}}^{(t)}(\sqrt{\varpi_{d,\bar{m}}},\psi_{d,\bar{m}})\geq  e^{\frac{\chi_{\text{UL}}}{\bar{M}}}-1$, \label{eq:MDDCF:SINRdbarmEqu:a} \\
&$0<\varpi_{d,\bar{m}} \leq p_{d\bar{m}}L^2$, \label{eq:MDDGNN:SINRdbarmEqu:b} \\
&$\psi_{d,m} \geq \sum_{l \in \mathcal{L}}\upsilon_{ld\bar{m}}\left(\xi_l^{\text{SI}}\Theta_{\text{UL}}+\sigma^2\right)$. \label{eq:MDDGNN:SINRdbarmEqu:c}
\end{subnumcases}

\begin{algorithm}
\caption{QT-SCA Algorithm for SE maximization in MDD-CF} 
\label{MDDGNN:al2}
\textbf{Initialization:} \\
Compute $\left\{\omega_{ldm}\right\}, \left\{\upsilon_{ld\bar{m}}\right\}, \forall l, d, m, \bar{m}$\;
Set $t=0$, $\varpi_{d,m}^{(0)}=1, \varpi_{d,\bar{m}}^{(0)}=1, \psi_{d,m}^{(0)}=1, \psi_{d,\bar{m}}^{(0)}=1, \forall d, m, \bar{m}$\;
Compute $\pmb{p}^{(0)}$ by solving optimization problem \eqref{eq:MDDGNN:pini};\\
\QT{}{
\Repeat{$\textup{Convergence}$}{
Compute $\pmb{z}^{(t)}$ using $\pmb{z}^{(t)}=\frac{\sqrt{A(\pmb{p}^{(t)})}}{B(\pmb{p}^{(t)})}$ for a fixed $\pmb{p}^{(t)}$ \;
Update $\pmb{p}^{(t+1)}$ via \eqref{eq:MDDGNN:SE_reformulation}, for a fixed $\pmb{z}^{(t)}$  \;
Update $\varpi_{d,m}^{(t+1)}, \varpi_{d,\bar{m}}^{(t+1)}, \psi_{d,m}^{(t+1)}, \psi_{d,\bar{m}}^{(t+1)}$\;
Set $t = t + 1$;
}
}

\KwOut{$\pmb{p}$}
\end{algorithm}

To this point, the optimization problem \eqref{eq:MDDGNN:SE_formulation} can be rewritten as  
\begin{align}\label{eq:MDDGNN:SE_reformulation}
\max_{\pmb{p},\pmb{z}, \pmb{\varpi}, \pmb{\psi}} &\frac{1}{M_{\text{sum}}}\sum_{d=1}^D \big(\sum_{m=1}^M \ln\big(1+2z_{dm}\sqrt{A_{d,m}(\pmb{p})}-z_{dm}^2B_{d,m}(\pmb{p})\big) \nonumber \\
&+\sum_{\bar{m}=1}^{\bar{M}} \ln(1+2z_{d\bar{m}}\sqrt{A_{d,\bar{m}}(\pmb{p})}-z_{d\bar{m}}^2 B_{d,\bar{m}}(\pmb{p})\big) \nonumber \\ 
\text{s.t.} \ &z_{dm} \in \mathbb{R}, \ \forall d \in \mathcal{D}, m \in \mathcal{M}, \nonumber \\ 
&z_{d\bar{m}} \in \mathbb{R}, \ \forall d \in \mathcal{D}, \bar{m} \in \mathcal{\bar{M}},\nonumber \\ 
&(\ref{eq:MDDGNN:SE_formulation}\text{b}), (\ref{eq:MDDGNN:SE_formulation}\text{c}), \eqref{eq:MDDGNN:SINRdmEqu:b}, \eqref{eq:MDDGNN:SINRdmEqu:c}, \eqref{eq:MDDGNN:SINRCA}, \eqref{eq:MDDGNN:SINRdbarmEqu}.
\end{align}
During the optimization of \eqref{eq:MDDGNN:SE_reformulation}, $\pmb{p}$ and $\pmb{z}$ are iteratively optimized. Hence, we first need to choose the feasible values of $\pmb{p}$. Specifically, $(\varpi_{d,m}^{(0)},\psi_{d,m}^{(0)})$ and $(\varpi_{d,\bar{m}}^{(0)},\psi_{d,\bar{m}}^{(0)})$ can be set to 1, for $\forall d \in \mathcal{D}, m \in \mathcal{M}, \bar{m} \in \mathcal{\bar{M}}$. With this regard, we resort to the optimization problem described as
\begin{align}\label{eq:MDDGNN:pini}
&\argmax_{\pmb{p}, \pmb{\varpi}, \pmb{\psi}} \sum_{d=1}^D\left(\sum_{m=1}^{M}\alpha_{d,m}+\sum_{\bar{m}=1}^{\bar{M}}\alpha_{d,\bar{m}}\right) \nonumber \\
\text{s.t.} \ \  &\alpha_{d,m}\leq 0, \alpha_{d,\bar{m}}\leq 0, \forall d \in \mathcal{D}, m \in \mathcal{M}, \bar{m} \in \mathcal{\bar{M}}, \nonumber \\
& \ 2\varpi_{d,m}-\psi_{d,m} \geq e^{\frac{\chi_{\text{DL}}}{M}}-1 + \alpha_{d,m}, \forall d \in \mathcal{D}, m \in \mathcal{M}, \nonumber \\
& \ 2\varpi_{d,\bar{m}}-\psi_{d,\bar{m}} \geq e^{\frac{\chi_{\text{UL}}}{\bar{M}}}-1 + \alpha_{d,\bar{m}}, \forall d \in \mathcal{D}, \bar{m} \in \mathcal{\bar{M}}, \nonumber \\
&~~(\ref{eq:MDDGNN:SE_formulation}\text{b}), (\ref{eq:MDDGNN:SE_formulation}\text{c}), \eqref{eq:MDDGNN:SINRdmEqu:b}, \eqref{eq:MDDGNN:SINRdmEqu:c}, \eqref{eq:MDDGNN:SINRdbarmEqu:b}, \eqref{eq:MDDGNN:SINRdbarmEqu:c},
\end{align}
where $(\left\{\alpha_{d,m}\right\}, \left\{\alpha_{d,\bar{m}}\right\})$ are new variables, and when they are zeros, the initial $\pmb{p}$ can be appropriately obtained. 

The overall QT-SCA algorithm is summarized as Algorithm \ref{MDDGNN:al2}. Note that, as the proposed QT-SCA algorithm mainly relies on the QT and SCA methods, the detailed proof of their convergence can be found in \cite{shen2018fractional} and \cite{marks1978general}, respectively. Additionally, the convergence behavior and computational complexity of the algorithm are evaluated in Section \ref{sec:MDDGNN:sim} of this paper.

\ifCLASSOPTIONcaptionsoff
  \newpage
\fi



%\bibliographystyle{IEEEtran}

\bibliographystyle{IEEEtran}
\bibliography{MDDRef}

\begin{thebibliography}{10}
\providecommand{\url}[1]{#1}
\csname url@samestyle\endcsname
\providecommand{\newblock}{\relax}
\providecommand{\bibinfo}[2]{#2}
\providecommand{\BIBentrySTDinterwordspacing}{\spaceskip=0pt\relax}
\providecommand{\BIBentryALTinterwordstretchfactor}{4}
\providecommand{\BIBentryALTinterwordspacing}{\spaceskip=\fontdimen2\font plus
\BIBentryALTinterwordstretchfactor\fontdimen3\font minus
  \fontdimen4\font\relax}
\providecommand{\BIBforeignlanguage}[2]{{%
\expandafter\ifx\csname l@#1\endcsname\relax
\typeout{** WARNING: IEEEtran.bst: No hyphenation pattern has been}%
\typeout{** loaded for the language `#1'. Using the pattern for}%
\typeout{** the default language instead.}%
\else
\language=\csname l@#1\endcsname
\fi
#2}}
\providecommand{\BIBdecl}{\relax}
\BIBdecl

\bibitem{demir2021foundations}
{\"O}.~T. Demir, E.~Bj{\"o}rnson, and L.~Sanguinetti, ``Foundations of
  user-centric cell-free massive {MIMO},'' \emph{arXiv preprint
  arXiv:2108.02541}, 2021.

\bibitem{goyal2015full}
S.~Goyal, P.~Liu, S.~S. Panwar, R.~A. Difazio, R.~Yang, and E.~Bala, ``Full
  duplex cellular systems: will doubling interference prevent doubling
  capacity?'' \emph{IEEE Communications Magazine}, vol.~53, no.~5, pp.
  121--127, 2015.

\bibitem{elhoushy2021cell}
S.~Elhoushy, M.~Ibrahim, and W.~Hamouda, ``Cell-free massive {{MIMO}}: A
  survey,'' \emph{IEEE Communications Surveys \& Tutorials}, 2021.

\bibitem{kolodziej2019band}
K.~E. Kolodziej, B.~T. Perry, and J.~S. Herd, ``In-band full-duplex technology:
  Techniques and systems survey,'' \emph{IEEE Transactions on Microwave Theory
  and Techniques}, vol.~67, no.~7, pp. 3025--3041, 2019.

\bibitem{everett2014passive}
E.~Everett, A.~Sahai, and A.~Sabharwal, ``Passive self-interference suppression
  for full-duplex infrastructure nodes,'' \emph{IEEE Transactions on Wireless
  Communications}, vol.~13, no.~2, pp. 680--694, 2014.

\bibitem{quan2017two}
X.~Quan, Y.~Liu, W.~Pan, Y.~Tang, and K.~Kang, ``A two-stage analog
  cancellation architecture for self-interference suppression in full-duplex
  communications,'' in \emph{2017 IEEE MTT-S International Microwave Symposium
  (IMS)}.\hskip 1em plus 0.5em minus 0.4em\relax IEEE, 2017, pp. 1169--1172.

\bibitem{ahmed2015all}
E.~Ahmed and A.~M. Eltawil, ``All-digital self-interference cancellation
  technique for full-duplex systems,'' \emph{IEEE Transactions on Wireless
  Communications}, vol.~14, no.~7, pp. 3519--3532, 2015.

\bibitem{liang2015digital}
D.~Liang, P.~Xiao, G.~Chen, M.~Ghoraishi, and R.~Tafazolli, ``Digital
  self-interference cancellation for full-duplex {{MIMO}} systems,'' in
  \emph{2015 International Wireless Communications and Mobile Computing
  Conference (IWCMC)}.\hskip 1em plus 0.5em minus 0.4em\relax IEEE, 2015, pp.
  403--407.

\bibitem{ghoraishi2015subband}
M.~Ghoraishi, W.~Jiang, P.~Xiao, and R.~Tafazolli, ``Subband approach for
  wideband self-interference cancellation in full-duplex transceiver,'' in
  \emph{2015 International Wireless Communications and Mobile Computing
  Conference (IWCMC)}.\hskip 1em plus 0.5em minus 0.4em\relax IEEE, 2015, pp.
  1139--1143.

\bibitem{li2021multicarrier}
B.~Li, L.-L. Yang, R.~G. Maunder, P.~Xiao, and S.~Sun, ``Multicarrier-division
  duplex: A duplexing technique for the shift to {{6G}} wireless
  communications,'' \emph{IEEE Vehicular Technology Magazine}, 2021.

\bibitem{da2021full}
J.~M.~B. da~Silva, G.~Wikstr{\"o}m, R.~K. Mungara, and C.~Fischione, ``Full
  duplex and dynamic {TDD}: Pushing the limits of spectrum reuse in multi-cell
  communications,'' \emph{IEEE Wireless Communications}, vol.~28, no.~1, pp.
  44--50, 2021.

\bibitem{kim2020dynamic}
H.~Kim, J.~Kim, and D.~Hong, ``Dynamic {TDD} systems for {5G} and beyond: A
  survey of cross-link interference mitigation,'' \emph{IEEE Communications
  Surveys \& Tutorials}, vol.~22, no.~4, pp. 2315--2348, 2020.

\bibitem{nguyen2020spectral}
H.~V. Nguyen, V.-D. Nguyen, O.~A. Dobre, S.~K. Sharma, S.~Chatzinotas,
  B.~Ottersten, and O.-S. Shin, ``On the spectral and energy efficiencies of
  full-duplex cell-free massive {MIMO},'' \emph{IEEE Journal on Selected Areas
  in Communications}, vol.~38, no.~8, pp. 1698--1718, 2020.

\bibitem{xia2021joint}
X.~Xia, P.~Zhu, J.~Li, H.~Wu, D.~Wang, Y.~Xin, and X.~You, ``Joint user
  selection and transceiver design for cell-free with network-assisted full
  duplexing,'' \emph{IEEE Transactions on Wireless Communications}, 2021.

\bibitem{wang2019performance}
D.~Wang, M.~Wang, P.~Zhu, J.~Li, J.~Wang, and X.~You, ``Performance of
  network-assisted full-duplex for cell-free massive {MIMO},'' \emph{IEEE
  Transactions on Communications}, vol.~68, no.~3, pp. 1464--1478, 2019.

\bibitem{li2022Spectral}
B.~Li, L.-L. Yang, R.~Maunder, S.~Sun, and P.~Xiao, ``Spectral-efficiency of
  cell-free massive {{MIMO}} with multicarrier-division duplex,'' \emph{arXiv
  preprint arXiv:2206.08774}, 2022.

\bibitem{you2021energy}
L.~You, Y.~Huang, D.~Zhang, Z.~Chang, W.~Wang, and X.~Gao, ``Energy efficiency
  optimization for multi-cell massive {{MIMO}}: Centralized and distributed
  power allocation algorithms,'' \emph{IEEE Transactions on Communications},
  vol.~69, no.~8, pp. 5228--5242, 2021.

\bibitem{shen2018fractional}
K.~Shen and W.~Yu, ``Fractional programming for communication systems-part {I}:
  Power control and beamforming,'' \emph{IEEE Transactions on Signal
  Processing}, vol.~66, no.~10, pp. 2616--2630, 2018.

\bibitem{amudala2019spectral}
D.~N. Amudala, E.~Sharma, and R.~Budhiraja, ``Spectral efficiency optimization
  of spatially-correlated multi-pair full-duplex massive {{MIMO}} relaying,''
  \emph{IEEE Transactions on Communications}, vol.~67, no.~12, pp. 8346--8364,
  2019.

\bibitem{fang2019optimal}
F.~Fang, Z.~Ding, W.~Liang, and H.~Zhang, ``Optimal energy efficient power
  allocation with user fairness for uplink {{MC-NOMA}} systems,'' \emph{IEEE
  Wireless Communications Letters}, vol.~8, no.~4, pp. 1133--1136, 2019.

\bibitem{li2017optimal}
J.-W. Li and C.~Lin, ``On the optimal power allocation for two-way full-duplex
  af relay networks,'' \emph{IEEE Transactions on Signal Processing}, vol.~65,
  no.~21, pp. 5702--5715, 2017.

\bibitem{xu2016power}
X.~Xu, X.~Chen, M.~Zhao, S.~Zhou, C.-Y. Chi, and J.~Wang, ``Power-efficient
  distributed beamforming for full-duplex {{MIMO}} relaying networks,''
  \emph{IEEE Transactions on Vehicular Technology}, vol.~66, no.~2, pp.
  1087--1103, 2016.

\bibitem{lee2018deep}
W.~Lee, M.~Kim, and D.-H. Cho, ``Deep power control: Transmit power control
  scheme based on convolutional neural network,'' \emph{IEEE Communications
  Letters}, vol.~22, no.~6, pp. 1276--1279, 2018.

\bibitem{liang2019towards}
F.~Liang, C.~Shen, W.~Yu, and F.~Wu, ``Towards optimal power control via
  ensembling deep neural networks,'' \emph{IEEE Transactions on
  Communications}, vol.~68, no.~3, pp. 1760--1776, 2019.

\bibitem{luo2022downlink}
L.~Luo, J.~Zhang, S.~Chen, X.~Zhang, B.~Ai, and D.~W.~K. Ng, ``Downlink power
  control for cell-free massive {{MIMO}} with deep reinforcement learning,''
  \emph{IEEE Transactions on Vehicular Technology}, 2022.

\bibitem{zhao2021dynamic}
Y.~Zhao, I.~G. Niemegeers, and S.~M.~H. De~Groot, ``Dynamic power allocation
  for cell-free massive {{MIMO}}: Deep reinforcement learning methods,''
  \emph{IEEE Access}, vol.~9, pp. 102\,953--102\,965, 2021.

\bibitem{bashar2020deep}
M.~Bashar, A.~Akbari, K.~Cumanan, H.~Q. Ngo, A.~G. Burr, P.~Xiao, and
  M.~Debbah, ``Deep learning-aided finite-capacity fronthaul cell-free massive
  {{MIMO}} with zero forcing,'' in \emph{ICC 2020-2020 IEEE International
  Conference on Communications (ICC)}.\hskip 1em plus 0.5em minus 0.4em\relax
  IEEE, 2020, pp. 1--6.

\bibitem{zaher2021learning}
M.~Zaher, {\"O}.~T. Demir, E.~Bj{\"o}rnson, and M.~Petrova, ``Learning-based
  downlink power allocation in cell-free massive {{MIMO}} systems,''
  \emph{arXiv preprint arXiv:2109.03128}, 2021.

\bibitem{shen2020graph}
Y.~Shen, Y.~Shi, J.~Zhang, and K.~B. Letaief, ``Graph neural networks for
  scalable radio resource management: Architecture design and theoretical
  analysis,'' \emph{IEEE Journal on Selected Areas in Communications}, vol.~39,
  no.~1, pp. 101--115, 2020.

\bibitem{chowdhury2021unfolding}
A.~Chowdhury, G.~Verma, C.~Rao, A.~Swami, and S.~Segarra, ``Unfolding {{WMMSE}}
  using graph neural networks for efficient power allocation,'' \emph{IEEE
  Transactions on Wireless Communications}, vol.~20, no.~9, pp. 6004--6017,
  2021.

\bibitem{eisen2020optimal}
M.~Eisen and A.~Ribeiro, ``Optimal wireless resource allocation with random
  edge graph neural networks,'' \emph{IEEE Transactions on Signal Processing},
  vol.~68, pp. 2977--2991, 2020.

\bibitem{li2020self}
B.~Li, L.-L. Yang, R.~G. Maunder, and S.~Sun, ``Self-interference cancellation
  and channel estimation in multicarrier-division duplex systems with hybrid
  beamforming,'' \emph{IEEE Access}, vol.~8, pp. 160\,653--160\,669, 2020.

\bibitem{day2012full2}
B.~P. Day, A.~R. Margetts, D.~W. Bliss, and P.~Schniter, ``Full-duplex {{MIMO}}
  relaying: Achievable rates under limited dynamic range,'' \emph{IEEE Journal
  on Selected Areas in Communications}, vol.~30, no.~8, pp. 1541--1553, 2012.

\bibitem{ng2016power}
D.~W.~K. Ng, Y.~Wu, and R.~Schober, ``Power efficient resource allocation for
  full-duplex radio distributed antenna networks,'' \emph{IEEE Transactions on
  Wireless Communications}, vol.~15, no.~4, pp. 2896--2911, 2016.

\bibitem{bjornson2017massive}
E.~Bj{\"o}rnson, J.~Hoydis, L.~Sanguinetti \emph{et~al.}, ``Massive {{MIMO}}
  networks: Spectral, energy, and hardware efficiency,'' \emph{Foundations and
  Trends{\textregistered} in Signal Processing}, vol.~11, no. 3-4, pp.
  154--655, 2017.

\bibitem{jiang2011performance}
Y.~Jiang, M.~K. Varanasi, and J.~Li, ``Performance analysis of {ZF} and {MMSE}
  equalizers for {MIMO} systems: An in-depth study of the high {SNR} regime,''
  \emph{IEEE Transactions on Information Theory}, vol.~57, no.~4, pp.
  2008--2026, 2011.

\bibitem{sun2013mining}
Y.~Sun and J.~Han, ``Mining heterogeneous information networks: a structural
  analysis approach,'' \emph{Acm Sigkdd Explorations Newsletter}, vol.~14,
  no.~2, pp. 20--28, 2013.

\bibitem{wang2019heterogeneous}
X.~Wang, H.~Ji, C.~Shi, B.~Wang, Y.~Ye, P.~Cui, and P.~S. Yu, ``Heterogeneous
  graph attention network,'' in \emph{The world wide web conference}, 2019, pp.
  2022--2032.

\bibitem{cui2019spatial}
W.~Cui, K.~Shen, and W.~Yu, ``Spatial deep learning for wireless scheduling,''
  \emph{IEEE Journal on Selected Areas in Communications}, vol.~37, no.~6, pp.
  1248--1261, 2019.

\bibitem{velivckovic2017graph}
P.~Veli{\v{c}}kovi{\'c}, G.~Cucurull, A.~Casanova, A.~Romero, P.~Lio, and
  Y.~Bengio, ``Graph attention networks,'' \emph{arXiv preprint
  arXiv:1710.10903}, 2017.

\bibitem{fey2019fast}
M.~Fey and J.~E. Lenssen, ``Fast graph representation learning with pytorch
  geometric,'' \emph{arXiv preprint arXiv:1903.02428}, 2019.

\bibitem{kingma2014adam}
D.~P. Kingma and J.~Ba, ``Adam: A method for stochastic optimization,''
  \emph{arXiv preprint arXiv:1412.6980}, 2014.

\bibitem{cvx}
M.~Grant and S.~Boyd, ``{CVX}: Matlab software for disciplined convex
  programming, version 2.1,'' \url{http://cvxr.com/cvx}, Mar. 2014.

\bibitem{jang2003transmit}
J.~Jang and K.~B. Lee, ``Transmit power adaptation for multiuser {{OFDM}}
  systems,'' \emph{IEEE Journal on Selected Areas in Communications}, vol.~21,
  no.~2, pp. 171--178, 2003.

\bibitem{peaucelle2002user}
D.~Peaucelle, D.~Henrion, Y.~Labit, and K.~Taitz, ``User's guide for sedumi
  interface 1.04,'' \emph{LAAS-CNRS, Toulouse}, 2002.

\bibitem{marks1978general}
B.~R. Marks and G.~P. Wright, ``A general inner approximation algorithm for
  nonconvex mathematical programs,'' \emph{Operations research}, vol.~26,
  no.~4, pp. 681--683, 1978.

\end{thebibliography}


%\end{IEEEbiography}


%\newpage

%%\begin{IEEEbiographynophoto}{Jane Doe}


%\vfill


\end{document}